\documentclass[twocolumn,showpacs,preprintnumbers,amsmath,amssymb]{revtex4-1}
\usepackage{graphicx} 
\usepackage{dcolumn} 
\usepackage{bm} 
\usepackage{braket} 
\usepackage[usenames,dvipsnames]{color}
\usepackage[
colorlinks=true, citecolor=blue, urlcolor=blue, linkcolor=blue,
setpagesize=false, bookmarks=false
]{hyperref}

\usepackage[english]{babel}
\addto{\captionsenglish}{}

\usepackage[resetlabels]{multibib}
\newcites{SM}{Refs in Supplemental Material}

\begin{document}


\title{Photoinduced $\eta$ Pairing in the Hubbard Model}
\author{Tatsuya Kaneko$^{1}$, Tomonori Shirakawa$^{2,1,3,4}$,  Sandro Sorella$^{2,5,3}$, and Seiji Yunoki$^{1,3,4}$}
\affiliation{
$^1$Computational Condensed Matter Physics Laboratory, RIKEN Cluster for Pioneering Research (CPR), Wako, Saitama 351-0198, Japan\\
$^2$SISSA--International School for Advanced Studies, Via Bonomea 265, 34136 Trieste, Italy \\
$^3$Computational Materials Science Research Team, RIKEN Center for Computational Science (R-CCS), Kobe, Hyogo 650-0047, Japan \\
$^4$Computational Quantum Matter Research Team, RIKEN Center for Emergent Matter Science (CEMS), Wako, Saitama 351-0198, Japan\\
$^5$Democritos Simulation Center CNR--IOM Instituto Officina dei Materiali, Via Bonomea 265, 34136 Trieste, Italy
}
\date{\today}


\begin{abstract}  
By employing unbiased numerical methods, we show that pulse irradiation can induce 
unconventional superconductivity even in the Mott insulator of the Hubbard model.  
The superconductivity found here in the photoexcited state is due to the $\eta$-pairing mechanism, characterized  by 
staggered pair-density-wave oscillations in the off-diagonal long-range correlation, 
and is absent in the ground-state phase diagram; i.e.,  
it is induced neither by a change 
of the effective interaction of the Hubbard model nor by simple photocarrier doping.  
Because of the selection rule, we show that the nonlinear optical response is essential to increase 
the number of $\eta$ pairs and thus enhance the superconducting correlation in the photoexcited state. 
Our finding demonstrates that nonequilibrium many-body 
dynamics is an alternative pathway to access a new exotic quantum state 
that is absent in the ground-state phase diagram, 
and also provides an alternative mechanism for enhancing superconductivity.
\end{abstract}

\maketitle


Recent experiments have clearly demonstrated that nonequilibrium dynamics can induce many intriguing 
phenomena in condensed-matter materials~\cite{To06,IO06,YN08,HTEetal14,GCFetal16}. 
Among them, the most striking is the discovery of photoinduced 
transient superconducting behaviors in some  high-$T_{\rm c}$ cuprates~\cite{FTDetal11,HKNetal14,KHNetal14} 
and alkali-doped fullerenes~\cite{MCNetal16,CBJetal18}. 
It has also been theoretically shown that superconductivity can be enhanced or induced by pulse irradiation in 
models for these materials~\cite{SKGetal16,KWRetal17,IOI17,MG17}. 
In these studies, the main focus 
is a photoinduced  state with physical properties already present 
in the corresponding equilibrium phases.  
In the case of a Mott insulator (MI), photoinduced insulator-to-metal transitions have been reported in time-resolved experiments for several transition-metal 
and organic-molecular compounds~\cite{IOMetal03,OMWetal07,UMTetal08,OMKetal10,OMKetal11}. 
In the MI, the photoinduced metallic state has been recognized as a result of photocarrier doping by 
creating doublon-holon pairs with no peculiar electronic states emerging~\cite{OA08,Ok12,EW13}. 

In this Letter, we show that pulse irradiation can induce superconductivity 
even in the celebrated MI of the Hubbard model. 
The photoinduced superconductivity is due to the $\eta$-pairing mechanism, forming 
on-site singlet pairs that exhibit, unlike conventional $s$-wave superconductivity, 
the staggered off-diagonal long-range correlation with a phase of $\pi$. 
Because of the selection rule, the nonlinear optical response is essential to increasing the number of $\eta$ pairs,  
and thus enhancing the superconducting correlation. 
Therefore, our finding is distinct from the previous studies~\cite{RRBetal08,BBPetal13,KA16,SM} and  
provides an alternative mechanism for enhancing 
superconductivity via nonequilibrium dynamics.


To demonstrate that superconductivity can be photoinduced in 
a MI, here we consider 
the half-filled one-dimensional (1D) Hubbard model at zero temperature. 
However, our finding does not depend on spatial dimensionality~\cite{SM}. 
The model is described by the following Hamiltonian: 
\begin{equation}
{\hat {\mathcal{H}}} = - t_h \sum_{i,\sigma}  \left( {\hat c}_{i,\sigma}^{\dag}{\hat c}_{i+1,\sigma} + {\rm H.c.} \right)  
 + U \sum_{i}  {\hat n}_{i, \uparrow}  {\hat n}_{i, \downarrow},            
\label{PI-eta_eq1}                    
\end{equation}
where ${\hat c}_{i,\sigma}$ (${\hat c}_{i,\sigma}^{\dag}$) is the annihilation (creation) operator for an electron at site $i$  
with spin $\sigma$~($= \uparrow, \downarrow$) and ${\hat n}_{i, \sigma}={\hat c}_{i,\sigma}^{\dag}{\hat c}_{i,\sigma}$. 
$t_h$ is the hopping integral between the nearest-neighboring sites, while $U$ ($>0$) is the on-site 
repulsive interaction. 
At half-filling, the ground state (GS) of the repulsive 1D Hubbard model is the MI 
with strong antiferromagnetic correlations. 

A time-dependent external field is introduced via the Peierls phase 
in Eq.~(\ref{PI-eta_eq1}) by replacing 
$ t_h {\hat c}_{i,\sigma}^{\dag}{\hat c}_{i+1,\sigma} \rightarrow t_h e^{iA(t)} {\hat c}_{i,\sigma}^{\dag}{\hat c}_{i+1,\sigma}$~\cite{Pe33}, 
where 
$A(t)$ is the vector potential as a function of time $t$, and 
the light velocity $c$, the elementary charge $e$, 
the Planck constant $\hbar$, and the lattice constant are set to 1.  
We consider a pump pulse given as $A(t) = A_0 e^{-(t-t_0)^2/(2\sigma_p^2)} \cos \left[ \omega_p (t-t_0)  \right]$
with the amplitude $A_0$, the frequency $\omega_p$, and the pulse width $\sigma_p$ centered at time  
$t_0\, (>0)$~\cite{TIA08,FCNetal12,LSMetal12,HI16,WCMetal17}.
With finite $A(t)$, the Hamiltonian becomes time dependent, 
$\hat{\mathcal{H}} \rightarrow \hat{\mathcal{H}}(t)$, 
and the equilibrium GS of $\hat{\cal{H}}$ at $t=0$ 
evolves in time, indicated here by $|\Psi(t)\rangle$.  
We employ the time-dependent exact diagonalization 
(ED) method for a finite-size cluster of $L$ (even) sites with periodic boundary conditions (PBC) 
to solve the time-dependent Schr\"odinger equation~\cite{SM}.  
We set $t_h$ ($t_h^{-1}$) as a unit of energy (time) 
and the total number $N$ of electrons to be $L$ at half-filling.


\begin{figure}[t]
\begin{center}
\includegraphics[width=\columnwidth]{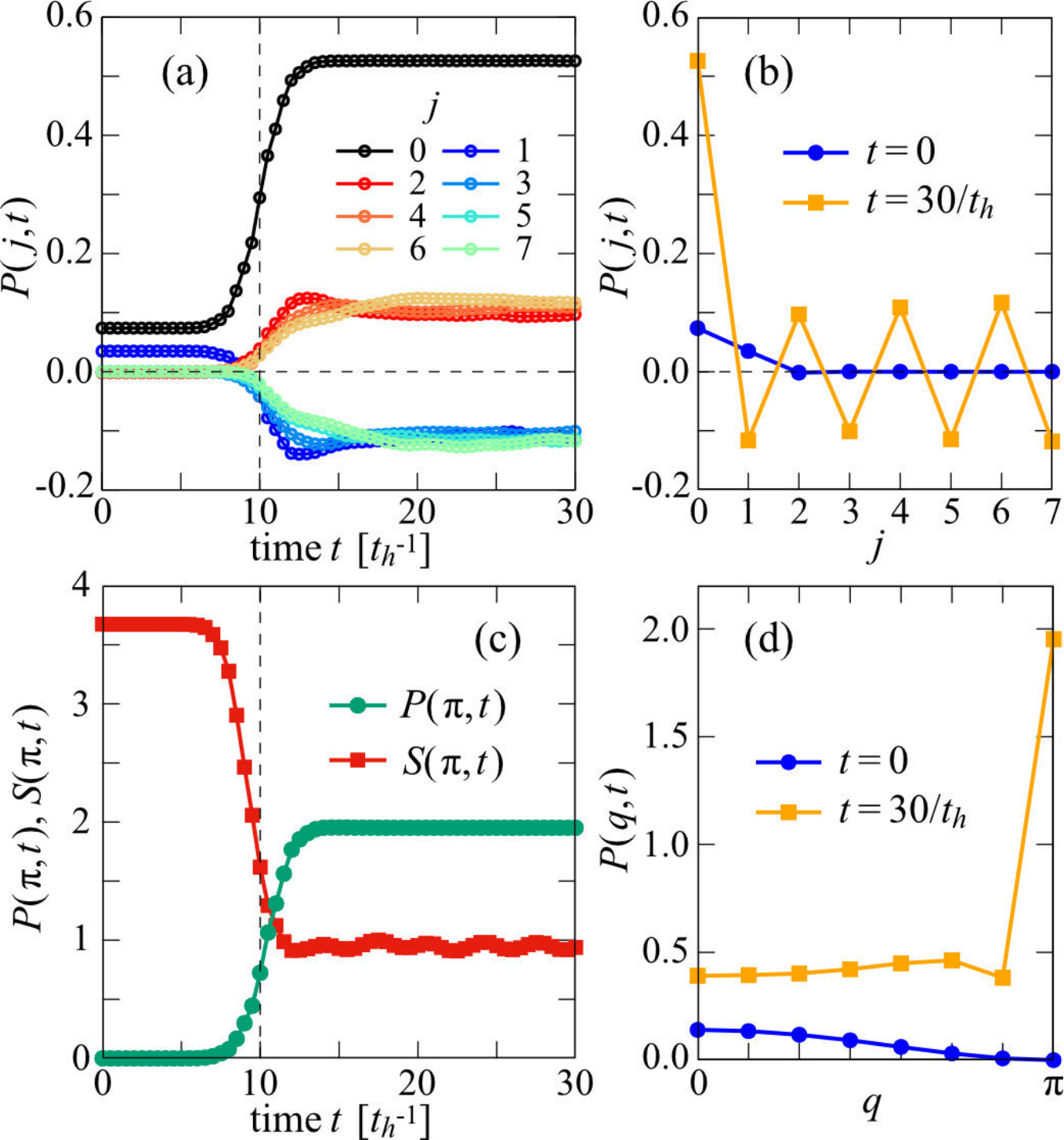}
\caption{
(a)~Time evolution of the on-site 
pair-correlation function $P(j,t)$. 
(b)~$P(j,t)$ at $t=0$ and $30/t_h$. 
(c)~Time evolution of the pair structure factor $P(q,t)$ and the spin structure factor 
$S(q,t)$ at $q=\pi$. 
(d)~$P(q,t)$ at $t=0$ and $30/t_h$. 
The results are calculated by the ED method for $L=14$ at $U=8t_h$ with 
$A_0=0.4$, $\omega_p=8.2t_h$, $\sigma_p=2/t_h$, and $t_0=10/t_h$. 
}
\label{fig:pair_corr}
\end{center}
\end{figure}

Enhancement of the double occupancy 
$n_d(t) = \frac{1}{L}\sum_{i} \bra{\Psi(t)} {\hat n}_{i,\uparrow} {\hat n}_{i,\downarrow} \ket{\Psi(t)}$ 
has been already reported in photoexcited states of the MIs~\cite{EW11,WHE14,YTY15,HI16}. 
Here, we find a significant increase of the superconducting pair correlation for the on-site singlet pair 
$\hat{\Delta}_i = \hat{c}_{i,\uparrow}\hat{c}_{i,\downarrow}$ after the pulse irradiation. 
Figure~\ref{fig:pair_corr}(a) shows the time evolution of the real-space pair-correlation function defined as 
$P(j,t)  = \frac{1}{L}\sum_{i}  \bra{\Psi(t)} \left( \hat{\Delta}^{\dag}_{i+j} \hat{\Delta}_{i}  + {\rm H.c.} \right) \ket{\Psi(t)}$.
Notice that $P(j,t)$ at $j=0$ corresponds to the double occupancy, i.e., $P(j \! = \! 0,t) = 2 n_d (t)$. 
We thus confirm the enhancement of $n_d (t)$ by the pulse irradiation. 
Surprisingly, $P(j \! \ne \! 0,t)$ is also enhanced significantly by the pulse irradiation and oscillates with the opposite 
phases between odd and even sites. 

As shown in Fig.~\ref{fig:pair_corr}(b), the pair correlation after the pulse irradiation extends to longer distances 
over the cluster, while the pair correlation is essentially absent in the initial MI state before the pulse irradiation.  
It is also clear that the sign of $P(j,t)$ alternates between neighboring sites, similar to a density wave, 
and accordingly the pair structure factor $P(q,t) =\sum_j e^{iqR_j} P(j,t) $, where $R_j$ is the location of site $j$, 
shows a sharp peak at $q=\pi$ [see Fig.~\ref{fig:pair_corr}(d)]. 
The time evolution of $P(q,t)$ and the spin structure factor $S(q,t) =\sum_j e^{iqR_j} S(j,t)$, 
where $S(j,t)=\frac{1}{L}\sum_{i}  \bra{\Psi(t)} \hat{m}^z_{i+j} \hat{m}^z_i \ket{\Psi(t)}$ 
and $\hat{m}^z_i = \hat{n}_{i,\uparrow}-\hat{n}_{i,\downarrow}$, 
is also calculated at $q=\pi$ in Fig.~\ref{fig:pair_corr}(c).
The antiferromagnetic correlation $S(q=\pi,t)$ is suppressed by the pulse irradiation, 
while the pair correlation 
$P(q=\pi,t)$ is strongly enhanced despite the fact that it is exactly zero before the pulse irradiation. 
Our matrix product state calculations also find 
the large enhancement of the pair correlation even for larger clusters 
that cannot be treated by the ED method~\cite{SM}.  

\begin{figure}[t]
\begin{center}
\includegraphics[width=0.9\columnwidth]{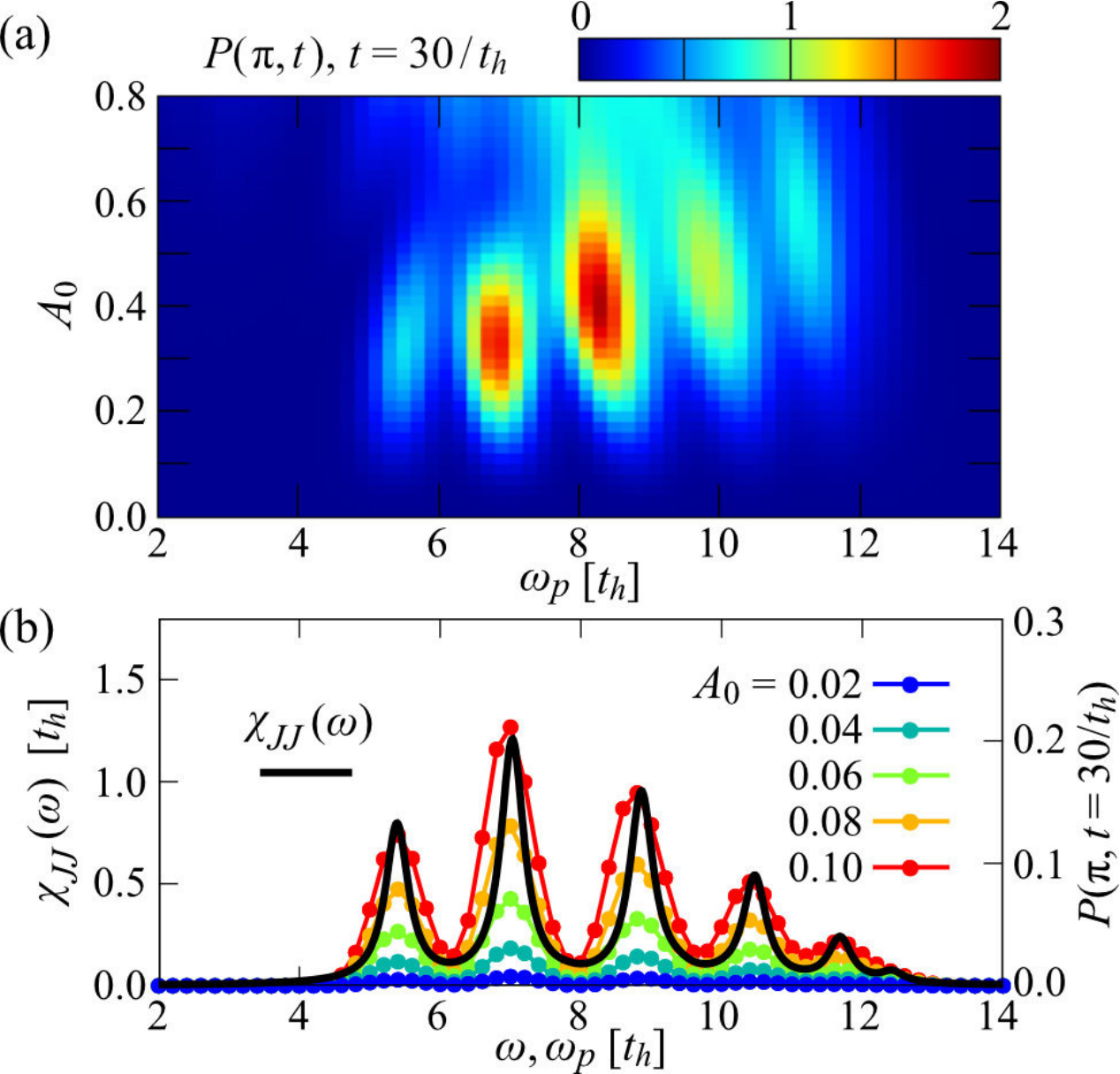}
\caption{
(a)~Contour plot of the pair structure factor $P(q=\pi,t)$ at $t=30/t_h$ with varying $\omega_p$ 
and $A_0$. 
(b)~The GS optical spectrum $\chi_{JJ}(\omega)$ is compared with $P(q=\pi,t=30/t_h)$ as a function of 
$\omega_p$ for different values of $A_0$. 
The results are calculated by the ED method for $L=14$ at $U=8t_h$, with $\sigma_p=2/t_h$ 
and $t_0=10/t_h$. 
}
\label{fig:optimal}
\end{center}
\end{figure}

In order to identify the optimal control parameters for the enhancement of $P(q=\pi,t)$, Fig.~\ref{fig:optimal}(a) 
shows the contour plot of $P(\pi,t)$ 
after the pulse irradiation with different values of $A_0$ and $\omega_p$.
For small $A_0$, we find that the peak structure of $P(q=\pi,t)$ as a function of $\omega_p$ is essentially the same  
as the GS optical spectrum 
$\chi_{JJ}(\omega) = \frac{1}{L}\bra{\psi_0} \hat{J} \delta(\omega-\hat{\mathcal{H}}+E_0) \hat{J} \ket{\psi_0}$, 
where $\ket{\psi_0}$ is the GS of $\hat{\cal{H}}$ with its energy $E_0$ and 
$\hat{J}=it_h\sum_{i,\sigma} (\hat{c}^{\dag}_{i+1,\sigma} \hat{c}_{i,\sigma} - \hat{c}^{\dag}_{i,\sigma} \hat{c}_{i+1,\sigma})$ is the current operator [see Fig.~\ref{fig:optimal}(b)]. 
This agreement is highly nontrivial and the reason will be clear below. 
$P(q=\pi,t)$ after the pulse irradiation is the largest at $A_0\sim0.4$ and $\omega_p \sim 8t_h$ $(=U)$. 
We should emphasize that the enhancement of $P(q=\pi,t)$ cannot be explained simply by the photodoping of carriers 
into the MI or due to a dynamical phase transition induced by effectively varying the model parameters, 
because there is no region in the GS phase diagram 
of the Hubbard model showing large on-site pairing correlations.


Instead, the behavior of the on-site pairs in the photoinduced state shown in Fig.~\ref{fig:pair_corr} 
can be understood in terms of  the so-called  
$\eta$-pairing, a concept originally introduced by Yang~\cite{Ya89}. 
In order to define the $\eta$-pairing, let us first introduce the following operators: 
$\hat{\eta}^+_j=(-1)^j \hat{c}^{\dag}_{j,\downarrow} \hat{c}^{\dag}_{j,\uparrow}$, 
$\hat{\eta}^-_j=(-1)^j \hat{c}_{j,\uparrow} \hat{c}_{j,\downarrow}, $ and 
$\hat{\eta}^z_j=\frac{1}{2}\left( {\hat n}_{j, \uparrow} + {\hat n}_{j, \downarrow} -1 \right)$. 
Notice that $\hat{\eta}^+_j$ ($\hat{\eta}^-_j$) is the same as $\hat{\Delta}_j^{\dag}$ ($\hat{\Delta}_j$)
except for the phase factor. 
These operators satisfy the $SU(2)$ commutation relations, i.e., $[ \hat{\eta}^+_j, \hat{\eta}^-_j] = 2 \hat{\eta}^z_j$ and $[ \hat{\eta}^z_j,  \hat{\eta}^\pm_j ] = \pm \hat{\eta}^\pm_j$. 
Similarly, the total $\hat{\eta}$ operators, 
$\hat{\eta}^\pm=\sum_j\hat{\eta}^\pm_j$ and $\hat{\eta}_z=\sum_j\hat{\eta}^z_j$,  
satisfy the $SU(2)$ commutation relations.  
The essential property of the $\hat{\eta}$ operators here is 
that they also satisfy 
$[\hat{\mathcal{H}},\hat{\eta}^{\pm}] = \pm U \hat{\eta}^{\pm}$ 
with $\hat{\mathcal{H}}$ in Eq.~(\ref{PI-eta_eq1}).

Yang originally proposed the $\eta$-pairing state $\ket{\phi_{N_{\eta}}} \propto (\hat{\eta}^{+})^{N_{\eta}} \ket{0}$, 
where $\ket{0}$ is a vacuum with no electrons and $N_{\eta}$ is the number of $\eta$ pairs~\cite{Ya89}.
Yang's $\eta$-pairing state $\ket{\phi_{N_{\eta}}}$ has two remarkable properties~\cite{Ya89}:  
First, $\ket{\phi_{N_{\eta}}}$ is an exact eigenstate of the Hubbard model with $2N_\eta$ electrons, 
satisfying $\hat{\mathcal{H}} \ket{\phi_{N_{\eta}}} = N_{\eta} U \ket{\phi_{N_{\eta}}}$. 
Second, $\bra{\phi_{N_{\eta}}} \hat{\Delta}^\dag_i\hat{\Delta}_j \ket{\phi_{N_{\eta}}} = \frac{N_{\eta}(L-N_{\eta})}{L(L-1)} 
e^{i\pi(R_i-R_j)}$ for $i\ne j$, 
indicating that $\ket{\phi_{N_{\eta}}}$ exhibits off-diagonal long-range order. 
Notice that both Yang's $\eta$-pairing state $\ket{\phi_{N_{\eta}}}$ and our photoinduced state $\ket{\Psi(t)}$ 
show similar sign-alternating characters in the pair-correlation function. 
However, the photoinduced state $\ket{\Psi(t)}$ excited from the MI state is different from the $\eta$-pairing state 
$\ket{\phi_{N_{\eta}}}$,  
in which all electrons participate in forming $\eta$ pairs, 
because we find numerically that $|\braket{\phi_{N_{\eta}}|\Psi(t)}|^2 = 0$ at $t=30/t_h$. 

As a candidate of the photoinduced state showing large $P(q \! = \! \pi,t)$, 
we now consider the eigenstate generated from the lowest-weight state (LWS) for 
$\hat{\eta}$ operators. 
For this purpose, it is important to notice that 
$[ \hat{\cal{H}}, \hat{\eta}^+\hat{\eta}^- ] = [   \hat{\cal{H}}, \hat{\eta}_z ]=0$. Therefore, 
any eigenstate of 
$\hat{\cal{H}}$ is also the eigenstate $|\eta, \eta_z\rangle $ of 
${\hat{\eta}}^2$ and $\hat{\eta}_z$ with the eigenvalues $\eta(\eta+1)$ and $\eta_z$, respectively, 
where ${\hat{\eta}}^2=\frac{1}{2}\left( \hat{\eta}^+\hat{\eta}^- + \hat{\eta}^-\hat{\eta}^+  \right) + \hat{\eta}_z^2$, 
$\eta=0,1,2,\cdots,\frac{L}{2}$ (at half-filling with the same number of up and down electrons 
$N_\uparrow=N_\downarrow$), and $\eta_z=-\eta,-\eta+1,\cdots,\eta$. 
This is precisely the analogue to the total spin operator $\hat{S}$ and its $z$ component $\hat{S}_z$ 
characterizing any eigenstate of 
$\hat{\cal{H}}$ with $| S, S_z \rangle_{\rm spin}$. The LWS is $|\eta,\eta_z=-\eta\rangle$ and thus satisfies 
$\hat{\eta}^- |\eta,-\eta\rangle =0$. 
Remarkably, Essler {\it et al.} have shown analytically that all the regular Bethe ansatz eigenstates of the 1D 
Hubbard model are the LWSs, and the remaining eigenstates can be generated from 
the LWSs by applying $\hat{\eta}^+$~\cite{EKS91,EKS92,EFGetal05}. 

Following them, we can construct the eigenstate having $N_{\eta}$ $\eta$ pairs from the LWS 
with $N_{\uparrow} = N_{\downarrow} = N_0$ ($\le L/2)$ as 
$\ket{\psi_{N_{\eta}}} = \frac{1}{\sqrt{\mathcal{C}_{N_{\eta}}}} (\hat{\eta}^+)^{N_{\eta}} \ket{\eta=\frac{L}{2}-N_0,\eta_z=-\eta}$~\cite{NL}. 
Yang's $\eta$-pairing state $\ket{\phi_{N_{\eta}}}$ corresponds to $\ket{\psi_{N_{\eta}}}$ 
generated from the vacuum state with $N_0=0$. 
At half-filling, $\ket{\psi_{N_{\eta}}}$ should contain $L$ electrons, and thus 
we consider $\ket{\psi_{N_{\eta}}}$ with 
$N_0 = L/2 - N_{\eta}$. Therefore, in this case, 
$\ket{\psi_{N_{\eta}}}\propto \ket{\eta=N_\eta,\eta_z=0}$, and hence 
$\bra{\psi_{N_{\eta}}} \hat{\eta}^+\hat{\eta}^- \ket{\psi_{N_{\eta}}} = N_\eta(N_\eta+1)$. 

\begin{figure}[t]
\begin{center}
\includegraphics[width=\columnwidth]{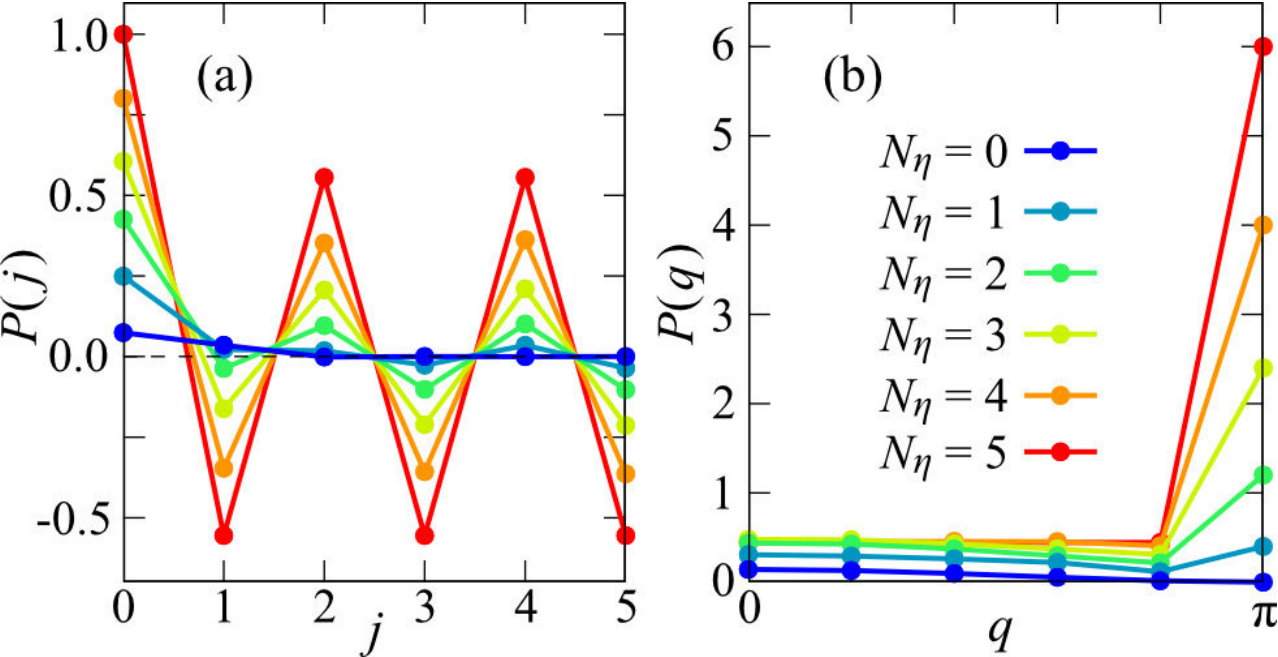}
\caption{
(a) On-site pair-correlation function $P(j)$ and (b) structure factor $P(q)$ 
for $\ket{\psi_{N_{\eta}}}$ 
at $U=8t_h$ with the different number of $\eta$ pairs $N_{\eta}$ ($\le L/2$).
$\ket{\psi_{N_{\eta}}}$ is generated from the ground state $\ket{\psi^{({\rm GS})}_{N_0,N_0}}$ of 
$\hat{\cal{H}}$ with $N_0 = L/2-N_{\eta}$ calculated by the ED for $L=10$ under PBC. 
}
\label{fig:SC_LWS}
\end{center}
\end{figure}

As an example, we construct $\ket{\psi_{N_{\eta}}}$ from the ground state 
$\ket{\psi^{({\rm GS})}_{N_\uparrow,N_\downarrow}}$ of $\hat{\cal{H}}$ 
with $N_{\uparrow} = N_{\downarrow} = N_0$~\cite{EEE}, which is the LWS. 
Figure~\ref{fig:SC_LWS} shows the on-site pair correlation, $P(j)$ and $P(q)$, 
for $\ket{\psi_{N_{\eta}}}$ 
with different $N_{\eta}$'s generated from $\ket{\psi^{({\rm GS})}_{N_0,N_0}}$. 
The sign-alternating character in $P(j)$ and the enhancement of $P(q\!=\! \pi)$ are clearly observed. 
This is understood because 
$P(q=\pi) = 2\bra{\psi_{N_{\eta}}}  \hat{\eta}^+\hat{\eta}^- \ket{\psi_{N_{\eta}}}/L = 2 N_{\eta}(N_{\eta} +1)/L$. 
With increasing $N_{\eta}$, $\ket{\psi_{N_{\eta}}}$ crossovers to Yang's 
$\eta$-pairing state $\ket{\phi_{N_{\eta}=L/2}}$ at $N_{\eta}=L/2$, for which $P(q=\pi)$ is the largest. 

\begin{figure*}[t]
\begin{center}
\includegraphics[width=0.87\textwidth]{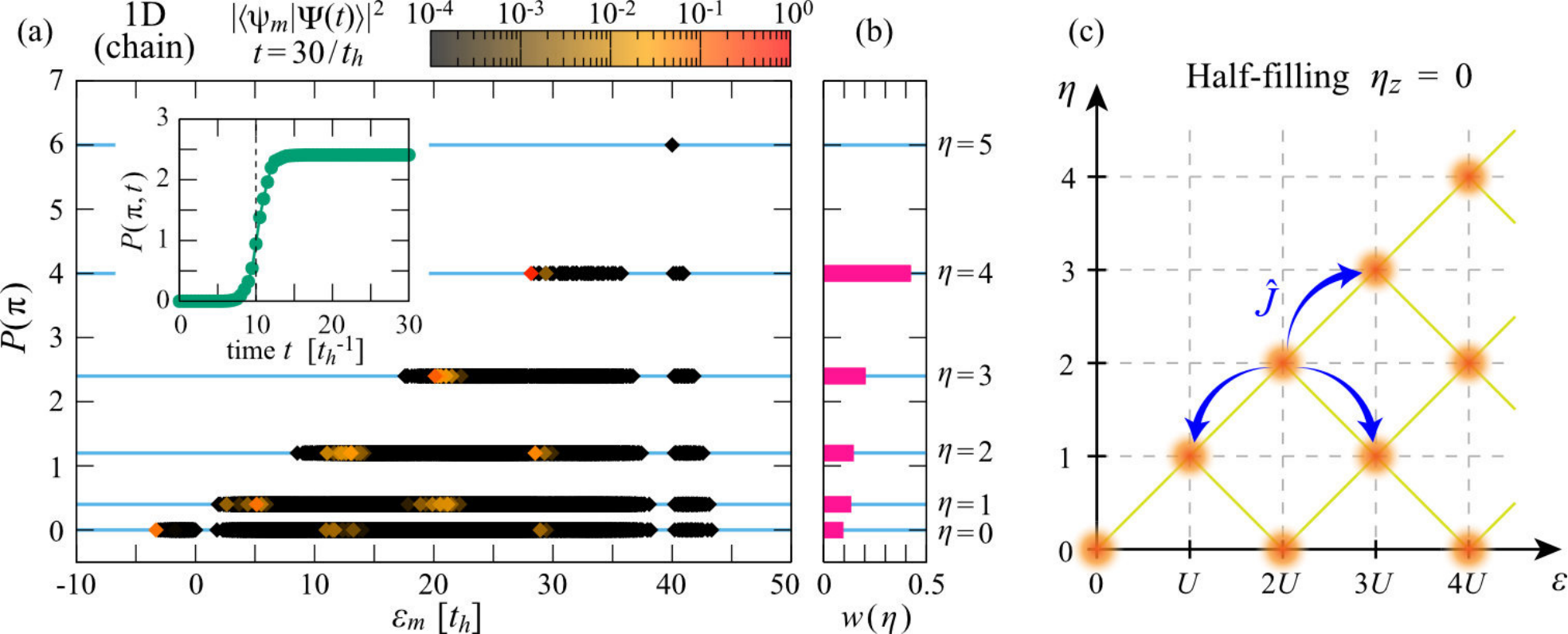}
\caption{
(a)~All eigenenergies $\varepsilon_m$ and $P(q=\pi)$ for the eigenstates $\ket{\psi_m}$ of 
the half-filled Hubbard Hamiltonian 
$\hat{\cal{H}}$ at $U=8t_h$ and $L=10$ under PBC. 
The color of each point indicates the weight $|\braket{\psi_m|\Psi(t)}|^2$ of the eigenstate $\ket{\psi_m}$ 
in the photoinduced state $\ket{\Psi(t)}$ at $t=30/t_h$ for $A(t)$ with $A_0=0.4$, 
$\omega_p=7.8t_h$, $\sigma_p=2/t_h$, 
and  $t_0=10/t_h$. 
The inset shows the time evolution of $P(q=\pi,t)$ for $\ket{\Psi(t)}$. 
(b) The total weight $w(\eta)$ of $|\braket{\psi_m|\Psi(t)}|^2$ over the states $\ket{\psi_m}$ with the same number 
$\eta$ of $\eta$ pairs in (a). Note that $\sum_{\eta=0}^{L/2}w(\eta)=1$. 
(c) Schematic figure of a ``tower of states" $\ket{\psi_m}$ in the photoinduced state  $\ket{\Psi(t)}$. 
The initial state before the pulse irradiation is at $(\varepsilon,\eta)=(0,0)$. 
The current operator $\hat{J}$ can induce the transition between states with $\Delta\eta=\pm1$ and 
$\Delta \varepsilon\sim\pm U$, 
as indicated by arrows, assuming that $\omega_p\sim U$, 
and thus the pulse irradiation eventually excites a series of 
states with nonzero $\eta$ and $\varepsilon$ (indicated by orange spheres). 
}
\label{fig:structure}
\end{center}
\end{figure*}

To elucidate the nature of the photoinduced state $\ket{\Psi(t)}$ in terms of the $\eta$ pairs, 
we calculate the eigenenergies $\varepsilon_m$ and the structure factors $P(q\!=\!\pi)$ 
for all the eigenstates $\ket{\psi_m}$ of $\hat{\cal{H}}$ at half-filling. 
As shown in Fig.~\ref{fig:structure}(a), the structure factor $P(q\!=\!\pi)$ for each eigenstate is nicely quantized. 
This is because each eigenstate $\ket{\psi_m}$ is also the eigenstate of $\hat{\eta}^2$ and $\hat{\eta}_z$, 
and the quantized values are given as $P(q=\pi) = 2\bra{\psi_{m}}\hat{\eta}^+\hat{\eta}^- \ket{\psi_{m}}/L 
= 2 \eta(\eta +1)/L$, with $\eta=0,1,\cdots,\frac{L}{2}$, corresponding to 
the number of $\eta$ pairs. 
These quantized values are exactly the same as $P(q=\pi) $ calculated for $\ket{\psi_{N_{\eta}}}$ 
in Fig.~\ref{fig:SC_LWS}(b).  

In Fig.~\ref{fig:structure}(a), the color of each point indicates the weight $|\braket{\psi_m|\Psi(t)}|^2$ of the eigenstate $\ket{\psi_m}$ 
in the photoinduced state $\ket{\Psi(t)}$
that exhibits the strong enhancement of $P(q=\pi,t)$ after the pulse irradiation [see the inset of Fig.~\ref{fig:structure}(a)].
We find that the state $\ket{\Psi(t)}$ after the pulse irradiation contains the nonzero weights of the eigenstates 
$\ket{\psi_m}$ with finite $\eta$ [also see Fig.~\ref{fig:structure}(b)]. 
This is exactly the reason for the photoinduced enhancement of $P(q=\pi,t)$. 
The Hubbard model itself has the eigenstates with $P(q\!=\!\pi)\ne0$, and the photoinduced state 
$\ket{\Psi(t)}$ captures the weights of those eigenstates. 

The process of the enhancement of $P(\pi,t)$ is explained as follows: 
Before the pulse irradiation, the initial state is the GS of 
$\hat{\cal H}$ with $\ket{\eta=0,\eta_z=0}$, i.e., the $\eta$ singlet state~\cite{EFGetal05}, and $P(q=\pi) = 0$. 
The pulse irradiation via $A(t)$ breaks the commutation relation as
$[\hat{\mathcal{H}}(t),\hat{\eta}^+] = [\hat{\mathcal{H}},\hat{\eta}^+] + \sum_{k} F(k,t) \hat{c}^{\dag}_{\pi-k,\downarrow} \hat{c}^{\dag}_{k ,\uparrow}$, with $F(k,t) = 4t_h \sin[A(t)] \sin k$, 
and this transient breaking of the $\eta$ symmetry stirs states with different values of $\eta$. 
After the pulse irradiation, the Hamiltonian again satisfies the commutation relation because $A(t) = 0$, 
but $\ket{\Psi(t)}$ now contains components of $\ket{\eta\ne0,\eta_z=0}$, which enhance $P(\pi,t)$.

More precisely, in the small-$A_0$ limit, the external perturbation is expressed as $A(t)\hat{J}$. 
We can show that the current operator $\hat{J}$ is a rank-one tensor operator with the zeroth component 
in terms of the $\hat{\eta}$ operators~\cite{SM}. Therefore, 
according to the Wigner-Eckart theorem~\cite{JJSakurai,*MERose}, 
there exists the selection rule such that 
$\bra{\eta',\eta_z'}\hat{J}\ket{\eta,\eta_z}\ne0$ only for $\eta'=\eta\pm1$ when $\eta'_z=\eta_z=0$ at half-filling. 
This implies that in the linear response regime the photoinduced state $\ket{\Psi(t)}$ can contain 
the eigenstates $\ket{\psi_m}$ with $\eta=1$ and the eigenenergies $\varepsilon_m\sim U$, assuming that 
$\omega_p$ is tuned around $U$. 
This explains the good agreement between the optical spectrum $\chi_{JJ}(\omega)$ and $P(q=\pi,t)$ 
found in Fig.~\ref{fig:optimal}(b). 
At the second order, the photoinduced state $\ket{\Psi(t)}$ can contain 
eigenstates $\ket{\psi_m}$ of $\hat{\mathcal{H}}$ with $\eta=2$ and $\varepsilon_m\sim 2U$, as well as 
$\eta=0$ and $\varepsilon_m\sim 0$ and $2U$. Applying the same argument for higher orders, 
$\eta$-pairing eigenstates with even larger $\eta$ values acquire in the transient period 
a finite overlap $|\braket{\psi_m|\Psi(t)}|^2$ with the photoinduced state. 
Considering all orders, eventually, the distribution of eigenstates $\ket{\psi_m}$ in the photoinduced state  
$\ket{\Psi(t)}$ forms a ``tower of states" shown schematically in Fig.~\ref{fig:structure}(c), 
which is indeed in good qualitative accordance with the numerical results in Fig.~\ref{fig:structure}(a) 
(for the analysis in the limit of $\sigma_p\to\infty$, see the Supplemental Material~\cite{SM}). 
This also explains why the pulse irradiation is effective to induce $\eta$ pairs, and 
the nonlinearity is essential to enhance the pair correlation. 
Note that the nonlinear response is absent in the noninteracting limit,  
clearly showing the importance of electron correlations. 
 
Exactly the same argument can be applied to the two-dimensional Hubbard model on the 
square lattice, and indeed we have found the large enhancement of the on-site pairing correlation in the photoinduced 
state, similar to the 1D case~\cite{SM}.
Although the enhancement of the pair correlation is 
most effective at half-filling, it remains even away from half-filling~\cite{SM}. 
We have also examined the effect of perturbation $\hat{\mathcal{H}}'$ 
that breaks the $\eta$ symmetry, i.e., $[\hat{\mathcal{H}}', \hat{\eta}^+\hat{\eta}^-]\ne0$, 
and still found the enhancement of the $\eta$-pairing correlation 
specially in the transient period~\cite{SM}.


In conclusion, we have found that density-wave-like staggered superconducting correlations are induced 
by photoexciting the MI ground state of the half-filled Hubbard model. 
The superconductivity is due to the $\eta$-pairing mechanism where the on-site singlet pairs display 
off-diagonal long-range correlation with phase $\pi$, the fingerprint 
of the $\eta$-pairing state. 
We have shown that the nonlinear optical response is essential to increase the number of $\eta$ pairs and hence 
enhance the superconducting correlation. 
The $\eta$-pairing states were originally introduced purely for the mathematical purpose to solve the Hubbard model 
analytically, and here we have demonstrated that the pulse irradiation can bring this object into the real world 
to be observed experimentally. 

Finally, we note that a more realistic treatment of materials should include a coupling with other degrees of freedom 
such as phonons, which introduces slow timescale dynamics in the thermalization process. 
Therefore, the $\eta$-pairing may be realized experimentally in a transient or prethermal regime. 
The most ideal system to explore the $\eta$-pairing experimentally is a cold 
fermionic atom system, for which the antiferromagnetic order has been recently observed~\cite{MCJetal17}.


The authors acknowledge S. Sota, K. Seki, S. Miyakoshi, T. Oka, S. Kitamura, P. Werner, Y. Murakami, and 
S. Ishihara for fruitful discussion. 
This work was supported in part by Grants-in-Aid for Scientific Research from MEXT Japan under Grants 
No.~JP17K05523, No.~JP18K13509, and No.~JP18H01183. 
T. S. acknowledges the Simons Foundation for financial support (Grant No.~534160).  
The authors are grateful for providing computational resources of the K computer in RIKEN R-CCS through the HPCI 
System Research Project (Projects  No.~hp140130, No.~hp150140, No.~hp170324, and No.~hp180098).
The calculations were also performed in part on the 
RIKEN supercomputer system (HOKUSAI GreatWave) at the Advanced Center for Computing and 
Communications (ACCC), RIKEN. 


\bibliography{TE-PI_Paper}

\begin{thebibliography}{15}%
\makeatletter
\providecommand \@ifxundefined [1]{%
 \@ifx{#1\undefined}
}%
\providecommand \@ifnum [1]{%
 \ifnum #1\expandafter \@firstoftwo
 \else \expandafter \@secondoftwo
 \fi
}%
\providecommand \@ifx [1]{%
 \ifx #1\expandafter \@firstoftwo
 \else \expandafter \@secondoftwo
 \fi
}%
\providecommand \natexlab [1]{#1}%
\providecommand \enquote  [1]{``#1''}%
\providecommand \bibnamefont  [1]{#1}%
\providecommand \bibfnamefont [1]{#1}%
\providecommand \citenamefont [1]{#1}%
\providecommand \href@noop [0]{\@secondoftwo}%
\providecommand \href [0]{\begingroup \@sanitize@url \@href}%
\providecommand \@href[1]{\@@startlink{#1}\@@href}%
\providecommand \@@href[1]{\endgroup#1\@@endlink}%
\providecommand \@sanitize@url [0]{\catcode `\\12\catcode `\$12\catcode
  `\&12\catcode `\#12\catcode `\^12\catcode `\_12\catcode `\%12\relax}%
\providecommand \@@startlink[1]{}%
\providecommand \@@endlink[0]{}%
\providecommand \url  [0]{\begingroup\@sanitize@url \@url }%
\providecommand \@url [1]{\endgroup\@href {#1}{\urlprefix }}%
\providecommand \urlprefix  [0]{URL }%
\providecommand \Eprint [0]{\href }%
\providecommand \doibase [0]{http://dx.doi.org/}%
\providecommand \selectlanguage [0]{\@gobble}%
\providecommand \bibinfo  [0]{\@secondoftwo}%
\providecommand \bibfield  [0]{\@secondoftwo}%
\providecommand \translation [1]{[#1]}%
\providecommand \BibitemOpen [0]{}%
\providecommand \bibitemStop [0]{}%
\providecommand \bibitemNoStop [0]{.\EOS\space}%
\providecommand \EOS [0]{\spacefactor3000\relax}%
\providecommand \BibitemShut  [1]{\csname bibitem#1\endcsname}%
\let\auto@bib@innerbib\@empty
\bibitem [{\citenamefont {Park}\ and\ \citenamefont {Light}(1986)}]{PL86S}%
  \BibitemOpen
  \bibfield  {author} {\bibinfo {author} {\bibfnamefont {T.~J.}\ \bibnamefont
  {Park}}\ and\ \bibinfo {author} {\bibfnamefont {J.}~\bibnamefont {Light}},\
  }\href {https://aip.scitation.org/doi/10.1063/1.451548} {\bibfield  {journal}
  {\bibinfo  {journal} {J. Chem. Phys.}\ }\textbf {\bibinfo {volume} {85}},\
  \bibinfo {pages} {5870} (\bibinfo {year} {1986})}\BibitemShut {NoStop}%
\bibitem [{\citenamefont {Mohankumar}\ and\ \citenamefont
  {Auerbach}(2006)}]{MA06S}%
  \BibitemOpen
  \bibfield  {author} {\bibinfo {author} {\bibfnamefont {N.}~\bibnamefont
  {Mohankumar}}\ and\ \bibinfo {author} {\bibfnamefont {S.~M.}\ \bibnamefont
  {Auerbach}},\ }\href {\doibase 10.1016/j.cpc.2006.07.005} {\bibfield
  {journal} {\bibinfo  {journal} {Comput. Phys. Commun.}\ }\textbf {\bibinfo
  {volume} {175}},\ \bibinfo {pages} {473} (\bibinfo {year}
  {2006})}\BibitemShut {NoStop}%
\bibitem [{\citenamefont {Hashimoto}\ and\ \citenamefont
  {Ishihara}(2016)}]{HI16S}%
  \BibitemOpen
  \bibfield  {author} {\bibinfo {author} {\bibfnamefont {H.}~\bibnamefont
  {Hashimoto}}\ and\ \bibinfo {author} {\bibfnamefont {S.}~\bibnamefont
  {Ishihara}},\ }\href {\doibase 10.1103/PhysRevB.93.165133} {\bibfield
  {journal} {\bibinfo  {journal} {Phys. Rev. B}\ }\textbf {\bibinfo {volume}
  {93}},\ \bibinfo {pages} {165133} (\bibinfo {year} {2016})}\BibitemShut
  {NoStop}%
\bibitem [{\citenamefont {P{\'e}rez-Garc{\'\i}a}\ \emph
  {et~al.}(2007)\citenamefont {P{\'e}rez-Garc{\'\i}a}, \citenamefont
  {Verstraete}, \citenamefont {Wolf},\ and\ \citenamefont
  {Cirac}}]{PVWetal07S}%
  \BibitemOpen
  \bibfield  {author} {\bibinfo {author} {\bibfnamefont {D.}~\bibnamefont
  {P{\'e}rez-Garc{\'\i}a}}, \bibinfo {author} {\bibfnamefont {F.}~\bibnamefont
  {Verstraete}}, \bibinfo {author} {\bibfnamefont {M.~M.}\ \bibnamefont
  {Wolf}}, \ and\ \bibinfo {author} {\bibfnamefont {J.~I.}\ \bibnamefont
  {Cirac}},\ }\href@noop {} {\bibfield  {journal} {\bibinfo  {journal} {Quantum
  Inf. Comput.}\ }\textbf {\bibinfo {volume} {7}},\ \bibinfo {pages} {401}
  (\bibinfo {year} {2007})}\BibitemShut {NoStop}%
\bibitem [{\citenamefont {White}(1992)}]{Wh92S}%
  \BibitemOpen
  \bibfield  {author} {\bibinfo {author} {\bibfnamefont {S.~R.}\ \bibnamefont
  {White}},\ }\href {\doibase 10.1103/PhysRevLett.69.2863} {\bibfield
  {journal} {\bibinfo  {journal} {Phys. Rev. Lett.}\ }\textbf {\bibinfo
  {volume} {69}},\ \bibinfo {pages} {2863} (\bibinfo {year}
  {1992})}\BibitemShut {NoStop}%
\bibitem [{\citenamefont {Schollw{\"o}ck}(2011)}]{Sc11S}%
  \BibitemOpen
  \bibfield  {author} {\bibinfo {author} {\bibfnamefont {U.}~\bibnamefont
  {Schollw{\"o}ck}},\ }\href {\doibase 10.1016/j.aop.2010.09.012} {\bibfield
  {journal} {\bibinfo  {journal} {Ann. Phys. (Amsterdam)}\ }\textbf {\bibinfo
  {volume} {326}},\ \bibinfo {pages} {96 } (\bibinfo {year}
  {2011})}\BibitemShut {NoStop}%
\bibitem [{\citenamefont {Zaletel}\ \emph {et~al.}(2015)\citenamefont
  {Zaletel}, \citenamefont {Mong}, \citenamefont {Karrasch}, \citenamefont
  {Moore},\ and\ \citenamefont {Pollmann}}]{ZMKetal15S}%
  \BibitemOpen
  \bibfield  {author} {\bibinfo {author} {\bibfnamefont {M.~P.}\ \bibnamefont
  {Zaletel}}, \bibinfo {author} {\bibfnamefont {R.~S.~K.}\ \bibnamefont
  {Mong}}, \bibinfo {author} {\bibfnamefont {C.}~\bibnamefont {Karrasch}},
  \bibinfo {author} {\bibfnamefont {J.~E.}\ \bibnamefont {Moore}}, \ and\
  \bibinfo {author} {\bibfnamefont {F.}~\bibnamefont {Pollmann}},\ }\href
  {\doibase 10.1103/PhysRevB.91.165112} {\bibfield  {journal} {\bibinfo
  {journal} {Phys. Rev. B}\ }\textbf {\bibinfo {volume} {91}},\ \bibinfo
  {pages} {165112} (\bibinfo {year} {2015})}\BibitemShut {NoStop}%
\bibitem [{ite()}]{itensorS}%
  \BibitemOpen
  \href@noop {} {}\bibinfo {note} {{http://itensor.org}}\BibitemShut {NoStop}%
\bibitem [{\citenamefont {Sakurai}(1994)}]{JJSakuraiS}%
  \BibitemOpen
  \bibfield  {author} {\bibinfo {author} {\bibfnamefont {J.~J.}\ \bibnamefont
  {Sakurai}},\ }\href@noop {} {\emph {\bibinfo {title} {Modern Quantum
  Mechanics}}}\ (\bibinfo  {publisher} {Addison-Wesley},\ \bibinfo {address}
  {Reading, MA},\ \bibinfo {year} {1994})\BibitemShut {NoStop}%
\bibitem [{\citenamefont {Rose}(1967)}]{MERoseS}%
  \BibitemOpen
  \bibfield  {author} {\bibinfo {author} {\bibfnamefont {M.~E.}\ \bibnamefont
  {Rose}},\ }\href@noop {} {\emph {\bibinfo {title} {Elementary Theory of
  Angular Momentum}}}\ (\bibinfo  {publisher} {Wiley},\ \bibinfo {address} {New
  York},\ \bibinfo {year} {1967})\BibitemShut {NoStop}%
\bibitem [{\citenamefont {Kitamura}\ and\ \citenamefont {Aoki}(2016)}]{KA16S}%
  \BibitemOpen
  \bibfield  {author} {\bibinfo {author} {\bibfnamefont {S.}~\bibnamefont
  {Kitamura}}\ and\ \bibinfo {author} {\bibfnamefont {H.}~\bibnamefont
  {Aoki}},\ }\href {\doibase 10.1103/PhysRevB.94.174503} {\bibfield  {journal}
  {\bibinfo  {journal} {Phys. Rev. B}\ }\textbf {\bibinfo {volume} {94}},\
  \bibinfo {pages} {174503} (\bibinfo {year} {2016})}\BibitemShut {NoStop}%
\bibitem [{\citenamefont {Rosch}\ \emph {et~al.}(2008)\citenamefont {Rosch},
  \citenamefont {Rasch}, \citenamefont {Binz},\ and\ \citenamefont
  {Vojta}}]{RRBetal08S}%
  \BibitemOpen
  \bibfield  {author} {\bibinfo {author} {\bibfnamefont {A.}~\bibnamefont
  {Rosch}}, \bibinfo {author} {\bibfnamefont {D.}~\bibnamefont {Rasch}},
  \bibinfo {author} {\bibfnamefont {B.}~\bibnamefont {Binz}}, \ and\ \bibinfo
  {author} {\bibfnamefont {M.}~\bibnamefont {Vojta}},\ }\href {\doibase
  10.1103/PhysRevLett.101.265301} {\bibfield  {journal} {\bibinfo  {journal}
  {Phys. Rev. Lett.}\ }\textbf {\bibinfo {volume} {101}},\ \bibinfo {pages}
  {265301} (\bibinfo {year} {2008})}\BibitemShut {NoStop}%
\bibitem [{\citenamefont {Bernier}\ \emph {et~al.}(2013)\citenamefont
  {Bernier}, \citenamefont {Barmettler}, \citenamefont {Poletti},\ and\
  \citenamefont {Kollath}}]{BBPetal13S}%
  \BibitemOpen
  \bibfield  {author} {\bibinfo {author} {\bibfnamefont {J.-S.}\ \bibnamefont
  {Bernier}}, \bibinfo {author} {\bibfnamefont {P.}~\bibnamefont {Barmettler}},
  \bibinfo {author} {\bibfnamefont {D.}~\bibnamefont {Poletti}}, \ and\
  \bibinfo {author} {\bibfnamefont {C.}~\bibnamefont {Kollath}},\ }\href
  {\doibase 10.1103/PhysRevA.87.063608} {\bibfield  {journal} {\bibinfo
  {journal} {Phys. Rev. A}\ }\textbf {\bibinfo {volume} {87}},\ \bibinfo
  {pages} {063608} (\bibinfo {year} {2013})}\BibitemShut {NoStop}%
\bibitem [{\citenamefont {Rojo}\ \emph {et~al.}(1990)\citenamefont {Rojo},
  \citenamefont {Sofo},\ and\ \citenamefont {Balseiro}}]{RSB90S}%
  \BibitemOpen
  \bibfield  {author} {\bibinfo {author} {\bibfnamefont {A.~G.}\ \bibnamefont
  {Rojo}}, \bibinfo {author} {\bibfnamefont {J.~O.}\ \bibnamefont {Sofo}}, \
  and\ \bibinfo {author} {\bibfnamefont {C.~A.}\ \bibnamefont {Balseiro}},\
  }\href {\doibase 10.1103/PhysRevB.42.10241} {\bibfield  {journal} {\bibinfo
  {journal} {Phys. Rev. B}\ }\textbf {\bibinfo {volume} {42}},\ \bibinfo
  {pages} {10241} (\bibinfo {year} {1990})}\BibitemShut {NoStop}%
\bibitem [{\citenamefont {Mentink}\ \emph {et~al.}(2015)\citenamefont
  {Mentink}, \citenamefont {Balzer},\ and\ \citenamefont {Eckstein}}]{MBE15S}%
  \BibitemOpen
  \bibfield  {author} {\bibinfo {author} {\bibfnamefont {J.}~\bibnamefont
  {Mentink}}, \bibinfo {author} {\bibfnamefont {K.}~\bibnamefont {Balzer}}, \
  and\ \bibinfo {author} {\bibfnamefont {M.}~\bibnamefont {Eckstein}},\ }\href
  {https://doi.org/10.1038/ncomms7708} {\bibfield  {journal} {\bibinfo
  {journal} {Nat. Commun.}\ }\textbf {\bibinfo {volume} {6}},\ \bibinfo {pages}
  {6708} (\bibinfo {year} {2015})}\BibitemShut {NoStop}%
\end{thebibliography}%


\begin{thebibliography}{54}%
\makeatletter
\providecommand \@ifxundefined [1]{%
 \@ifx{#1\undefined}
}%
\providecommand \@ifnum [1]{%
 \ifnum #1\expandafter \@firstoftwo
 \else \expandafter \@secondoftwo
 \fi
}%
\providecommand \@ifx [1]{%
 \ifx #1\expandafter \@firstoftwo
 \else \expandafter \@secondoftwo
 \fi
}%
\providecommand \natexlab [1]{#1}%
\providecommand \enquote  [1]{``#1''}%
\providecommand \bibnamefont  [1]{#1}%
\providecommand \bibfnamefont [1]{#1}%
\providecommand \citenamefont [1]{#1}%
\providecommand \href@noop [0]{\@secondoftwo}%
\providecommand \href [0]{\begingroup \@sanitize@url \@href}%
\providecommand \@href[1]{\@@startlink{#1}\@@href}%
\providecommand \@@href[1]{\endgroup#1\@@endlink}%
\providecommand \@sanitize@url [0]{\catcode `\\12\catcode `\$12\catcode
  `\&12\catcode `\#12\catcode `\^12\catcode `\_12\catcode `\%12\relax}%
\providecommand \@@startlink[1]{}%
\providecommand \@@endlink[0]{}%
\providecommand \url  [0]{\begingroup\@sanitize@url \@url }%
\providecommand \@url [1]{\endgroup\@href {#1}{\urlprefix }}%
\providecommand \urlprefix  [0]{URL }%
\providecommand \Eprint [0]{\href }%
\providecommand \doibase [0]{http://dx.doi.org/}%
\providecommand \selectlanguage [0]{\@gobble}%
\providecommand \bibinfo  [0]{\@secondoftwo}%
\providecommand \bibfield  [0]{\@secondoftwo}%
\providecommand \translation [1]{[#1]}%
\providecommand \BibitemOpen [0]{}%
\providecommand \bibitemStop [0]{}%
\providecommand \bibitemNoStop [0]{.\EOS\space}%
\providecommand \EOS [0]{\spacefactor3000\relax}%
\providecommand \BibitemShut  [1]{\csname bibitem#1\endcsname}%
\let\auto@bib@innerbib\@empty
\bibitem [{\citenamefont {Tokura}(2006)}]{To06}%
  \BibitemOpen
  \bibfield  {author} {\bibinfo {author} {\bibfnamefont {Y.}~\bibnamefont
  {Tokura}},\ }\href {\doibase 10.1143/JPSJ.75.011001} {\bibfield  {journal}
  {\bibinfo  {journal} {J. Phys. Soc. Jpn.}\ }\textbf {\bibinfo {volume}
  {75}},\ \bibinfo {pages} {011001} (\bibinfo {year} {2006})}\BibitemShut
  {NoStop}%
\bibitem [{\citenamefont {Iwai}\ and\ \citenamefont {Okamoto}(2006)}]{IO06}%
  \BibitemOpen
  \bibfield  {author} {\bibinfo {author} {\bibfnamefont {S.}~\bibnamefont
  {Iwai}}\ and\ \bibinfo {author} {\bibfnamefont {H.}~\bibnamefont {Okamoto}},\
  }\href {\doibase 10.1143/JPSJ.75.011007} {\bibfield  {journal} {\bibinfo
  {journal} {J. Phys. Soc. Jpn.}\ }\textbf {\bibinfo {volume} {75}},\ \bibinfo
  {pages} {011007} (\bibinfo {year} {2006})}\BibitemShut {NoStop}%
\bibitem [{\citenamefont {Yonemitsu}\ and\ \citenamefont {Nasu}(2008)}]{YN08}%
  \BibitemOpen
  \bibfield  {author} {\bibinfo {author} {\bibfnamefont {K.}~\bibnamefont
  {Yonemitsu}}\ and\ \bibinfo {author} {\bibfnamefont {K.}~\bibnamefont
  {Nasu}},\ }\href {\doibase 10.1016/j.physrep.2008.04.008} {\bibfield
  {journal} {\bibinfo  {journal} {Phys. Rep.}\ }\textbf {\bibinfo {volume}
  {465}},\ \bibinfo {pages} {1} (\bibinfo {year} {2008})}\BibitemShut {NoStop}%
\bibitem [{\citenamefont {Aoki}\ \emph {et~al.}(2014)\citenamefont {Aoki},
  \citenamefont {Tsuji}, \citenamefont {Eckstein}, \citenamefont {Kollar},
  \citenamefont {Oka},\ and\ \citenamefont {Werner}}]{HTEetal14}%
  \BibitemOpen
  \bibfield  {author} {\bibinfo {author} {\bibfnamefont {H.}~\bibnamefont
  {Aoki}}, \bibinfo {author} {\bibfnamefont {N.}~\bibnamefont {Tsuji}},
  \bibinfo {author} {\bibfnamefont {M.}~\bibnamefont {Eckstein}}, \bibinfo
  {author} {\bibfnamefont {M.}~\bibnamefont {Kollar}}, \bibinfo {author}
  {\bibfnamefont {T.}~\bibnamefont {Oka}}, \ and\ \bibinfo {author}
  {\bibfnamefont {P.}~\bibnamefont {Werner}},\ }\href {\doibase
  10.1103/RevModPhys.86.779} {\bibfield  {journal} {\bibinfo  {journal} {Rev.
  Mod. Phys.}\ }\textbf {\bibinfo {volume} {86}},\ \bibinfo {pages} {779}
  (\bibinfo {year} {2014})}\BibitemShut {NoStop}%
\bibitem [{\citenamefont {Giannetti}\ \emph {et~al.}(2016)\citenamefont
  {Giannetti}, \citenamefont {Capone}, \citenamefont {Fausti}, \citenamefont
  {Fabrizio}, \citenamefont {Parmigiani},\ and\ \citenamefont
  {Mihailovic}}]{GCFetal16}%
  \BibitemOpen
  \bibfield  {author} {\bibinfo {author} {\bibfnamefont {C.}~\bibnamefont
  {Giannetti}}, \bibinfo {author} {\bibfnamefont {M.}~\bibnamefont {Capone}},
  \bibinfo {author} {\bibfnamefont {D.}~\bibnamefont {Fausti}}, \bibinfo
  {author} {\bibfnamefont {M.}~\bibnamefont {Fabrizio}}, \bibinfo {author}
  {\bibfnamefont {F.}~\bibnamefont {Parmigiani}}, \ and\ \bibinfo {author}
  {\bibfnamefont {D.}~\bibnamefont {Mihailovic}},\ }\href {\doibase
  10.1080/00018732.2016.1194044} {\bibfield  {journal} {\bibinfo  {journal}
  {Adv. Phys.}\ }\textbf {\bibinfo {volume} {65}},\ \bibinfo {pages} {58}
  (\bibinfo {year} {2016})}\BibitemShut {NoStop}%
\bibitem [{\citenamefont {Fausti}\ \emph {et~al.}(2011)\citenamefont {Fausti},
  \citenamefont {Tobey}, \citenamefont {Dean}, \citenamefont {Kaiser},
  \citenamefont {Dienst}, \citenamefont {Hoffmann}, \citenamefont {Pyon},
  \citenamefont {Takayama}, \citenamefont {Takagi},\ and\ \citenamefont
  {Cavalleri}}]{FTDetal11}%
  \BibitemOpen
  \bibfield  {author} {\bibinfo {author} {\bibfnamefont {D.}~\bibnamefont
  {Fausti}}, \bibinfo {author} {\bibfnamefont {R.~I.}\ \bibnamefont {Tobey}},
  \bibinfo {author} {\bibfnamefont {N.}~\bibnamefont {Dean}}, \bibinfo {author}
  {\bibfnamefont {S.}~\bibnamefont {Kaiser}}, \bibinfo {author} {\bibfnamefont
  {A.}~\bibnamefont {Dienst}}, \bibinfo {author} {\bibfnamefont {M.~C.}\
  \bibnamefont {Hoffmann}}, \bibinfo {author} {\bibfnamefont {S.}~\bibnamefont
  {Pyon}}, \bibinfo {author} {\bibfnamefont {T.}~\bibnamefont {Takayama}},
  \bibinfo {author} {\bibfnamefont {H.}~\bibnamefont {Takagi}}, \ and\ \bibinfo
  {author} {\bibfnamefont {A.}~\bibnamefont {Cavalleri}},\ }\href {\doibase
  10.1126/science.1197294} {\bibfield  {journal} {\bibinfo  {journal}
  {Science}\ }\textbf {\bibinfo {volume} {331}},\ \bibinfo {pages} {189}
  (\bibinfo {year} {2011})}\BibitemShut {NoStop}%
\bibitem [{\citenamefont {Hu}\ \emph {et~al.}(2014)\citenamefont {Hu},
  \citenamefont {Kaiser}, \citenamefont {Nicoletti}, \citenamefont {Hunt},
  \citenamefont {Gierz}, \citenamefont {Hoffmann}, \citenamefont {{Le Tacon}},
  \citenamefont {Loew}, \citenamefont {Keimer},\ and\ \citenamefont
  {Cavalleri}}]{HKNetal14}%
  \BibitemOpen
  \bibfield  {author} {\bibinfo {author} {\bibfnamefont {W.}~\bibnamefont
  {Hu}}, \bibinfo {author} {\bibfnamefont {S.}~\bibnamefont {Kaiser}}, \bibinfo
  {author} {\bibfnamefont {D.}~\bibnamefont {Nicoletti}}, \bibinfo {author}
  {\bibfnamefont {C.~R.}\ \bibnamefont {Hunt}}, \bibinfo {author}
  {\bibfnamefont {I.}~\bibnamefont {Gierz}}, \bibinfo {author} {\bibfnamefont
  {M.~C.}\ \bibnamefont {Hoffmann}}, \bibinfo {author} {\bibfnamefont
  {M.}~\bibnamefont {{Le Tacon}}}, \bibinfo {author} {\bibfnamefont
  {T.}~\bibnamefont {Loew}}, \bibinfo {author} {\bibfnamefont {B.}~\bibnamefont
  {Keimer}}, \ and\ \bibinfo {author} {\bibfnamefont {A.}~\bibnamefont
  {Cavalleri}},\ }\href {\doibase 10.1038/nmat3963} {\bibfield  {journal}
  {\bibinfo  {journal} {Nat. Mater.}\ }\textbf {\bibinfo {volume} {13}},\
  \bibinfo {pages} {705} (\bibinfo {year} {2014})}\BibitemShut {NoStop}%
\bibitem [{\citenamefont {Kaiser}\ \emph {et~al.}(2014)\citenamefont {Kaiser},
  \citenamefont {Hunt}, \citenamefont {Nicoletti}, \citenamefont {Hu},
  \citenamefont {Gierz}, \citenamefont {Liu}, \citenamefont {Le~Tacon},
  \citenamefont {Loew}, \citenamefont {Haug}, \citenamefont {Keimer},\ and\
  \citenamefont {Cavalleri}}]{KHNetal14}%
  \BibitemOpen
  \bibfield  {author} {\bibinfo {author} {\bibfnamefont {S.}~\bibnamefont
  {Kaiser}}, \bibinfo {author} {\bibfnamefont {C.~R.}\ \bibnamefont {Hunt}},
  \bibinfo {author} {\bibfnamefont {D.}~\bibnamefont {Nicoletti}}, \bibinfo
  {author} {\bibfnamefont {W.}~\bibnamefont {Hu}}, \bibinfo {author}
  {\bibfnamefont {I.}~\bibnamefont {Gierz}}, \bibinfo {author} {\bibfnamefont
  {H.~Y.}\ \bibnamefont {Liu}}, \bibinfo {author} {\bibfnamefont
  {M.}~\bibnamefont {Le~Tacon}}, \bibinfo {author} {\bibfnamefont
  {T.}~\bibnamefont {Loew}}, \bibinfo {author} {\bibfnamefont {D.}~\bibnamefont
  {Haug}}, \bibinfo {author} {\bibfnamefont {B.}~\bibnamefont {Keimer}}, \ and\
  \bibinfo {author} {\bibfnamefont {A.}~\bibnamefont {Cavalleri}},\ }\href
  {\doibase 10.1103/PhysRevB.89.184516} {\bibfield  {journal} {\bibinfo
  {journal} {Phys. Rev. B}\ }\textbf {\bibinfo {volume} {89}},\ \bibinfo
  {pages} {184516} (\bibinfo {year} {2014})}\BibitemShut {NoStop}%
\bibitem [{\citenamefont {Mitrano}\ \emph {et~al.}(2016)\citenamefont
  {Mitrano}, \citenamefont {Cantaluppi}, \citenamefont {Nicoletti},
  \citenamefont {Kaiser}, \citenamefont {Perucchi}, \citenamefont {Lupi},
  \citenamefont {{Di Pietro}}, \citenamefont {Pontiroli}, \citenamefont
  {Ricc{\`{o}}}, \citenamefont {Clark}, \citenamefont {Jaksch},\ and\
  \citenamefont {Cavalleri}}]{MCNetal16}%
  \BibitemOpen
  \bibfield  {author} {\bibinfo {author} {\bibfnamefont {M.}~\bibnamefont
  {Mitrano}}, \bibinfo {author} {\bibfnamefont {A.}~\bibnamefont {Cantaluppi}},
  \bibinfo {author} {\bibfnamefont {D.}~\bibnamefont {Nicoletti}}, \bibinfo
  {author} {\bibfnamefont {S.}~\bibnamefont {Kaiser}}, \bibinfo {author}
  {\bibfnamefont {A.}~\bibnamefont {Perucchi}}, \bibinfo {author}
  {\bibfnamefont {S.}~\bibnamefont {Lupi}}, \bibinfo {author} {\bibfnamefont
  {P.}~\bibnamefont {{Di Pietro}}}, \bibinfo {author} {\bibfnamefont
  {D.}~\bibnamefont {Pontiroli}}, \bibinfo {author} {\bibfnamefont
  {M.}~\bibnamefont {Ricc{\`{o}}}}, \bibinfo {author} {\bibfnamefont {S.~R.}\
  \bibnamefont {Clark}}, \bibinfo {author} {\bibfnamefont {D.}~\bibnamefont
  {Jaksch}}, \ and\ \bibinfo {author} {\bibfnamefont {A.}~\bibnamefont
  {Cavalleri}},\ }\href {\doibase 10.1038/nature16522} {\bibfield  {journal}
  {\bibinfo  {journal} {Nature (London)}\ }\textbf {\bibinfo {volume} {530}},\
  \bibinfo {pages} {461} (\bibinfo {year} {2016})}\BibitemShut {NoStop}%
\bibitem [{\citenamefont {Cantaluppi}\ \emph {et~al.}(2018)\citenamefont
  {Cantaluppi}, \citenamefont {Buzzi}, \citenamefont {Jotzu}, \citenamefont
  {Nicoletti}, \citenamefont {Mitrano}, \citenamefont {Pontiroli},
  \citenamefont {Ricc{\`o}}, \citenamefont {Perucchi}, \citenamefont
  {Di~Pietro},\ and\ \citenamefont {Cavalleri}}]{CBJetal18}%
  \BibitemOpen
  \bibfield  {author} {\bibinfo {author} {\bibfnamefont {A.}~\bibnamefont
  {Cantaluppi}}, \bibinfo {author} {\bibfnamefont {M.}~\bibnamefont {Buzzi}},
  \bibinfo {author} {\bibfnamefont {G.}~\bibnamefont {Jotzu}}, \bibinfo
  {author} {\bibfnamefont {D.}~\bibnamefont {Nicoletti}}, \bibinfo {author}
  {\bibfnamefont {M.}~\bibnamefont {Mitrano}}, \bibinfo {author} {\bibfnamefont
  {D.}~\bibnamefont {Pontiroli}}, \bibinfo {author} {\bibfnamefont
  {M.}~\bibnamefont {Ricc{\`o}}}, \bibinfo {author} {\bibfnamefont
  {A.}~\bibnamefont {Perucchi}}, \bibinfo {author} {\bibfnamefont
  {P.}~\bibnamefont {Di~Pietro}}, \ and\ \bibinfo {author} {\bibfnamefont
  {A.}~\bibnamefont {Cavalleri}},\ }\href {\doibase 10.1038/s41567-018-0134-8}
  {\bibfield  {journal} {\bibinfo  {journal} {Nat. Phys.}\ }\textbf {\bibinfo
  {volume} {14}},\ \bibinfo {pages} {837} (\bibinfo {year} {2018})}\BibitemShut
  {NoStop}%
\bibitem [{\citenamefont {Sentef}\ \emph {et~al.}(2016)\citenamefont {Sentef},
  \citenamefont {Kemper}, \citenamefont {Georges},\ and\ \citenamefont
  {Kollath}}]{SKGetal16}%
  \BibitemOpen
  \bibfield  {author} {\bibinfo {author} {\bibfnamefont {M.~A.}\ \bibnamefont
  {Sentef}}, \bibinfo {author} {\bibfnamefont {A.~F.}\ \bibnamefont {Kemper}},
  \bibinfo {author} {\bibfnamefont {A.}~\bibnamefont {Georges}}, \ and\
  \bibinfo {author} {\bibfnamefont {C.}~\bibnamefont {Kollath}},\ }\href
  {\doibase 10.1103/PhysRevB.93.144506} {\bibfield  {journal} {\bibinfo
  {journal} {Phys. Rev. B}\ }\textbf {\bibinfo {volume} {93}},\ \bibinfo
  {pages} {144506} (\bibinfo {year} {2016})}\BibitemShut {NoStop}%
\bibitem [{\citenamefont {Kennes}\ \emph {et~al.}(2017)\citenamefont {Kennes},
  \citenamefont {Wilner}, \citenamefont {Reichman},\ and\ \citenamefont
  {Millis}}]{KWRetal17}%
  \BibitemOpen
  \bibfield  {author} {\bibinfo {author} {\bibfnamefont {D.~M.}\ \bibnamefont
  {Kennes}}, \bibinfo {author} {\bibfnamefont {E.~Y.}\ \bibnamefont {Wilner}},
  \bibinfo {author} {\bibfnamefont {D.~R.}\ \bibnamefont {Reichman}}, \ and\
  \bibinfo {author} {\bibfnamefont {A.~J.}\ \bibnamefont {Millis}},\ }\href
  {\doibase 10.1038/nphys4024} {\bibfield  {journal} {\bibinfo  {journal} {Nat.
  Phys.}\ }\textbf {\bibinfo {volume} {13}},\ \bibinfo {pages} {479} (\bibinfo
  {year} {2017})}\BibitemShut {NoStop}%
\bibitem [{\citenamefont {Ido}\ \emph {et~al.}(2017)\citenamefont {Ido},
  \citenamefont {Ohgoe},\ and\ \citenamefont {Imada}}]{IOI17}%
  \BibitemOpen
  \bibfield  {author} {\bibinfo {author} {\bibfnamefont {K.}~\bibnamefont
  {Ido}}, \bibinfo {author} {\bibfnamefont {T.}~\bibnamefont {Ohgoe}}, \ and\
  \bibinfo {author} {\bibfnamefont {M.}~\bibnamefont {Imada}},\ }\href
  {\doibase 10.1126/sciadv.1700718} {\bibfield  {journal} {\bibinfo  {journal}
  {Sci. Adv.}\ }\textbf {\bibinfo {volume} {3}},\ \bibinfo {pages} {e1700718}
  (\bibinfo {year} {2017})}\BibitemShut {NoStop}%
\bibitem [{\citenamefont {Mazza}\ and\ \citenamefont {Georges}(2017)}]{MG17}%
  \BibitemOpen
  \bibfield  {author} {\bibinfo {author} {\bibfnamefont {G.}~\bibnamefont
  {Mazza}}\ and\ \bibinfo {author} {\bibfnamefont {A.}~\bibnamefont
  {Georges}},\ }\href {\doibase 10.1103/PhysRevB.96.064515} {\bibfield
  {journal} {\bibinfo  {journal} {Phys. Rev. B}\ }\textbf {\bibinfo {volume}
  {96}},\ \bibinfo {pages} {064515} (\bibinfo {year} {2017})}\BibitemShut
  {NoStop}%
\bibitem [{\citenamefont {Iwai}\ \emph {et~al.}(2003)\citenamefont {Iwai},
  \citenamefont {Ono}, \citenamefont {Maeda}, \citenamefont {Matsuzaki},
  \citenamefont {Kishida}, \citenamefont {Okamoto},\ and\ \citenamefont
  {Tokura}}]{IOMetal03}%
  \BibitemOpen
  \bibfield  {author} {\bibinfo {author} {\bibfnamefont {S.}~\bibnamefont
  {Iwai}}, \bibinfo {author} {\bibfnamefont {M.}~\bibnamefont {Ono}}, \bibinfo
  {author} {\bibfnamefont {A.}~\bibnamefont {Maeda}}, \bibinfo {author}
  {\bibfnamefont {H.}~\bibnamefont {Matsuzaki}}, \bibinfo {author}
  {\bibfnamefont {H.}~\bibnamefont {Kishida}}, \bibinfo {author} {\bibfnamefont
  {H.}~\bibnamefont {Okamoto}}, \ and\ \bibinfo {author} {\bibfnamefont
  {Y.}~\bibnamefont {Tokura}},\ }\href {\doibase 10.1103/PhysRevLett.91.057401}
  {\bibfield  {journal} {\bibinfo  {journal} {Phys. Rev. Lett.}\ }\textbf
  {\bibinfo {volume} {91}},\ \bibinfo {pages} {057401} (\bibinfo {year}
  {2003})}\BibitemShut {NoStop}%
\bibitem [{\citenamefont {Okamoto}\ \emph {et~al.}(2007)\citenamefont
  {Okamoto}, \citenamefont {Matsuzaki}, \citenamefont {Wakabayashi},
  \citenamefont {Takahashi},\ and\ \citenamefont {Hasegawa}}]{OMWetal07}%
  \BibitemOpen
  \bibfield  {author} {\bibinfo {author} {\bibfnamefont {H.}~\bibnamefont
  {Okamoto}}, \bibinfo {author} {\bibfnamefont {H.}~\bibnamefont {Matsuzaki}},
  \bibinfo {author} {\bibfnamefont {T.}~\bibnamefont {Wakabayashi}}, \bibinfo
  {author} {\bibfnamefont {Y.}~\bibnamefont {Takahashi}}, \ and\ \bibinfo
  {author} {\bibfnamefont {T.}~\bibnamefont {Hasegawa}},\ }\href {\doibase
  10.1103/PhysRevLett.98.037401} {\bibfield  {journal} {\bibinfo  {journal}
  {Phys. Rev. Lett.}\ }\textbf {\bibinfo {volume} {98}},\ \bibinfo {pages}
  {037401} (\bibinfo {year} {2007})}\BibitemShut {NoStop}%
\bibitem [{\citenamefont {Uemura}\ \emph {et~al.}(2008)\citenamefont {Uemura},
  \citenamefont {Matsuzaki}, \citenamefont {Takahashi}, \citenamefont
  {Hasegawa},\ and\ \citenamefont {Okamoto}}]{UMTetal08}%
  \BibitemOpen
  \bibfield  {author} {\bibinfo {author} {\bibfnamefont {H.}~\bibnamefont
  {Uemura}}, \bibinfo {author} {\bibfnamefont {H.}~\bibnamefont {Matsuzaki}},
  \bibinfo {author} {\bibfnamefont {Y.}~\bibnamefont {Takahashi}}, \bibinfo
  {author} {\bibfnamefont {T.}~\bibnamefont {Hasegawa}}, \ and\ \bibinfo
  {author} {\bibfnamefont {H.}~\bibnamefont {Okamoto}},\ }\href {\doibase
  10.1143/JPSJ.77.113714} {\bibfield  {journal} {\bibinfo  {journal} {J. Phys.
  Soc. Jpn.}\ }\textbf {\bibinfo {volume} {77}},\ \bibinfo {pages} {113714}
  (\bibinfo {year} {2008})}\BibitemShut {NoStop}%
\bibitem [{\citenamefont {Okamoto}\ \emph {et~al.}(2010)\citenamefont
  {Okamoto}, \citenamefont {Miyagoe}, \citenamefont {Kobayashi}, \citenamefont
  {Uemura}, \citenamefont {Nishioka}, \citenamefont {Matsuzaki}, \citenamefont
  {Sawa},\ and\ \citenamefont {Tokura}}]{OMKetal10}%
  \BibitemOpen
  \bibfield  {author} {\bibinfo {author} {\bibfnamefont {H.}~\bibnamefont
  {Okamoto}}, \bibinfo {author} {\bibfnamefont {T.}~\bibnamefont {Miyagoe}},
  \bibinfo {author} {\bibfnamefont {K.}~\bibnamefont {Kobayashi}}, \bibinfo
  {author} {\bibfnamefont {H.}~\bibnamefont {Uemura}}, \bibinfo {author}
  {\bibfnamefont {H.}~\bibnamefont {Nishioka}}, \bibinfo {author}
  {\bibfnamefont {H.}~\bibnamefont {Matsuzaki}}, \bibinfo {author}
  {\bibfnamefont {A.}~\bibnamefont {Sawa}}, \ and\ \bibinfo {author}
  {\bibfnamefont {Y.}~\bibnamefont {Tokura}},\ }\href {\doibase
  10.1103/PhysRevB.82.060513} {\bibfield  {journal} {\bibinfo  {journal} {Phys.
  Rev. B}\ }\textbf {\bibinfo {volume} {82}},\ \bibinfo {pages} {060513}
  (\bibinfo {year} {2010})}\BibitemShut {NoStop}%
\bibitem [{\citenamefont {Okamoto}\ \emph {et~al.}(2011)\citenamefont
  {Okamoto}, \citenamefont {Miyagoe}, \citenamefont {Kobayashi}, \citenamefont
  {Uemura}, \citenamefont {Nishioka}, \citenamefont {Matsuzaki}, \citenamefont
  {Sawa},\ and\ \citenamefont {Tokura}}]{OMKetal11}%
  \BibitemOpen
  \bibfield  {author} {\bibinfo {author} {\bibfnamefont {H.}~\bibnamefont
  {Okamoto}}, \bibinfo {author} {\bibfnamefont {T.}~\bibnamefont {Miyagoe}},
  \bibinfo {author} {\bibfnamefont {K.}~\bibnamefont {Kobayashi}}, \bibinfo
  {author} {\bibfnamefont {H.}~\bibnamefont {Uemura}}, \bibinfo {author}
  {\bibfnamefont {H.}~\bibnamefont {Nishioka}}, \bibinfo {author}
  {\bibfnamefont {H.}~\bibnamefont {Matsuzaki}}, \bibinfo {author}
  {\bibfnamefont {A.}~\bibnamefont {Sawa}}, \ and\ \bibinfo {author}
  {\bibfnamefont {Y.}~\bibnamefont {Tokura}},\ }\href {\doibase
  10.1103/PhysRevB.83.125102} {\bibfield  {journal} {\bibinfo  {journal} {Phys.
  Rev. B}\ }\textbf {\bibinfo {volume} {83}},\ \bibinfo {pages} {125102}
  (\bibinfo {year} {2011})}\BibitemShut {NoStop}%
\bibitem [{\citenamefont {Oka}\ and\ \citenamefont {Aoki}(2008)}]{OA08}%
  \BibitemOpen
  \bibfield  {author} {\bibinfo {author} {\bibfnamefont {T.}~\bibnamefont
  {Oka}}\ and\ \bibinfo {author} {\bibfnamefont {H.}~\bibnamefont {Aoki}},\
  }\href {\doibase 10.1103/PhysRevB.78.241104} {\bibfield  {journal} {\bibinfo
  {journal} {Phys. Rev. B}\ }\textbf {\bibinfo {volume} {78}},\ \bibinfo
  {pages} {241104} (\bibinfo {year} {2008})}\BibitemShut {NoStop}%
\bibitem [{\citenamefont {Oka}(2012)}]{Ok12}%
  \BibitemOpen
  \bibfield  {author} {\bibinfo {author} {\bibfnamefont {T.}~\bibnamefont
  {Oka}},\ }\href {\doibase 10.1103/PhysRevB.86.075148} {\bibfield  {journal}
  {\bibinfo  {journal} {Phys. Rev. B}\ }\textbf {\bibinfo {volume} {86}},\
  \bibinfo {pages} {075148} (\bibinfo {year} {2012})}\BibitemShut {NoStop}%
\bibitem [{\citenamefont {Eckstein}\ and\ \citenamefont {Werner}(2013)}]{EW13}%
  \BibitemOpen
  \bibfield  {author} {\bibinfo {author} {\bibfnamefont {M.}~\bibnamefont
  {Eckstein}}\ and\ \bibinfo {author} {\bibfnamefont {P.}~\bibnamefont
  {Werner}},\ }\href {\doibase 10.1103/PhysRevLett.110.126401} {\bibfield
  {journal} {\bibinfo  {journal} {Phys. Rev. Lett.}\ }\textbf {\bibinfo
  {volume} {110}},\ \bibinfo {pages} {126401} (\bibinfo {year}
  {2013})}\BibitemShut {NoStop}%
\bibitem [{\citenamefont {Rosch}\ \emph {et~al.}(2008)\citenamefont {Rosch},
  \citenamefont {Rasch}, \citenamefont {Binz},\ and\ \citenamefont
  {Vojta}}]{RRBetal08}%
  \BibitemOpen
  \bibfield  {author} {\bibinfo {author} {\bibfnamefont {A.}~\bibnamefont
  {Rosch}}, \bibinfo {author} {\bibfnamefont {D.}~\bibnamefont {Rasch}},
  \bibinfo {author} {\bibfnamefont {B.}~\bibnamefont {Binz}}, \ and\ \bibinfo
  {author} {\bibfnamefont {M.}~\bibnamefont {Vojta}},\ }\href {\doibase
  10.1103/PhysRevLett.101.265301} {\bibfield  {journal} {\bibinfo  {journal}
  {Phys. Rev. Lett.}\ }\textbf {\bibinfo {volume} {101}},\ \bibinfo {pages}
  {265301} (\bibinfo {year} {2008})}\BibitemShut {NoStop}%
\bibitem [{\citenamefont {Bernier}\ \emph {et~al.}(2013)\citenamefont
  {Bernier}, \citenamefont {Barmettler}, \citenamefont {Poletti},\ and\
  \citenamefont {Kollath}}]{BBPetal13}%
  \BibitemOpen
  \bibfield  {author} {\bibinfo {author} {\bibfnamefont {J.-S.}\ \bibnamefont
  {Bernier}}, \bibinfo {author} {\bibfnamefont {P.}~\bibnamefont {Barmettler}},
  \bibinfo {author} {\bibfnamefont {D.}~\bibnamefont {Poletti}}, \ and\
  \bibinfo {author} {\bibfnamefont {C.}~\bibnamefont {Kollath}},\ }\href
  {\doibase 10.1103/PhysRevA.87.063608} {\bibfield  {journal} {\bibinfo
  {journal} {Phys. Rev. A}\ }\textbf {\bibinfo {volume} {87}},\ \bibinfo
  {pages} {063608} (\bibinfo {year} {2013})}\BibitemShut {NoStop}%
\bibitem [{\citenamefont {Kitamura}\ and\ \citenamefont {Aoki}(2016)}]{KA16}%
  \BibitemOpen
  \bibfield  {author} {\bibinfo {author} {\bibfnamefont {S.}~\bibnamefont
  {Kitamura}}\ and\ \bibinfo {author} {\bibfnamefont {H.}~\bibnamefont
  {Aoki}},\ }\href {\doibase 10.1103/PhysRevB.94.174503} {\bibfield  {journal}
  {\bibinfo  {journal} {Phys. Rev. B}\ }\textbf {\bibinfo {volume} {94}},\
  \bibinfo {pages} {174503} (\bibinfo {year} {2016})}\BibitemShut {NoStop}%
\bibitem [{SM()}]{SM}%
  \BibitemOpen
  \href@noop {} {}\bibinfo {note} {See Supplemental Material for details, which
  includes Refs.~\cite{PL86,MA06,PVWetal07,Wh92,
  Sc11,ZMKetal15,itensor,RSB90,MBE15}.}\BibitemShut {Stop}%
\bibitem [{\citenamefont {Park}\ and\ \citenamefont {Light}(1986)}]{PL86}%
  \BibitemOpen
  \bibfield  {author} {\bibinfo {author} {\bibfnamefont {T.~J.}\ \bibnamefont
  {Park}}\ and\ \bibinfo {author} {\bibfnamefont {J.}~\bibnamefont {Light}},\
  }\href {https://aip.scitation.org/doi/10.1063/1.451548} {\bibfield  {journal}
  {\bibinfo  {journal} {J. Chem. Phys.}\ }\textbf {\bibinfo {volume} {85}},\
  \bibinfo {pages} {5870} (\bibinfo {year} {1986})}\BibitemShut {NoStop}%
\bibitem [{\citenamefont {Mohankumar}\ and\ \citenamefont
  {Auerbach}(2006)}]{MA06}%
  \BibitemOpen
  \bibfield  {author} {\bibinfo {author} {\bibfnamefont {N.}~\bibnamefont
  {Mohankumar}}\ and\ \bibinfo {author} {\bibfnamefont {S.~M.}\ \bibnamefont
  {Auerbach}},\ }\href {\doibase 10.1016/j.cpc.2006.07.005} {\bibfield
  {journal} {\bibinfo  {journal} {Comput. Phys. Commun.}\ }\textbf {\bibinfo
  {volume} {175}},\ \bibinfo {pages} {473} (\bibinfo {year}
  {2006})}\BibitemShut {NoStop}%
\bibitem [{\citenamefont {P{\'e}rez-Garc{\'\i}a}\ \emph
  {et~al.}(2007)\citenamefont {P{\'e}rez-Garc{\'\i}a}, \citenamefont
  {Verstraete}, \citenamefont {Wolf},\ and\ \citenamefont {Cirac}}]{PVWetal07}%
  \BibitemOpen
  \bibfield  {author} {\bibinfo {author} {\bibfnamefont {D.}~\bibnamefont
  {P{\'e}rez-Garc{\'\i}a}}, \bibinfo {author} {\bibfnamefont {F.}~\bibnamefont
  {Verstraete}}, \bibinfo {author} {\bibfnamefont {M.~M.}\ \bibnamefont
  {Wolf}}, \ and\ \bibinfo {author} {\bibfnamefont {J.~I.}\ \bibnamefont
  {Cirac}},\ }\href@noop {} {\bibfield  {journal} {\bibinfo  {journal} {Quantum
  Inf. Comput.}\ }\textbf {\bibinfo {volume} {7}},\ \bibinfo {pages} {401}
  (\bibinfo {year} {2007})}\BibitemShut {NoStop}%
\bibitem [{\citenamefont {White}(1992)}]{Wh92}%
  \BibitemOpen
  \bibfield  {author} {\bibinfo {author} {\bibfnamefont {S.~R.}\ \bibnamefont
  {White}},\ }\href {\doibase 10.1103/PhysRevLett.69.2863} {\bibfield
  {journal} {\bibinfo  {journal} {Phys. Rev. Lett.}\ }\textbf {\bibinfo
  {volume} {69}},\ \bibinfo {pages} {2863} (\bibinfo {year}
  {1992})}\BibitemShut {NoStop}%
\bibitem [{\citenamefont {Schollw{\"o}ck}(2011)}]{Sc11}%
  \BibitemOpen
  \bibfield  {author} {\bibinfo {author} {\bibfnamefont {U.}~\bibnamefont
  {Schollw{\"o}ck}},\ }\href {\doibase 10.1016/j.aop.2010.09.012} {\bibfield
  {journal} {\bibinfo  {journal} {Ann. Phys. (Amsterdam)}\ }\textbf {\bibinfo
  {volume} {326}},\ \bibinfo {pages} {96 } (\bibinfo {year}
  {2011})}\BibitemShut {NoStop}%
\bibitem [{\citenamefont {Zaletel}\ \emph {et~al.}(2015)\citenamefont
  {Zaletel}, \citenamefont {Mong}, \citenamefont {Karrasch}, \citenamefont
  {Moore},\ and\ \citenamefont {Pollmann}}]{ZMKetal15}%
  \BibitemOpen
  \bibfield  {author} {\bibinfo {author} {\bibfnamefont {M.~P.}\ \bibnamefont
  {Zaletel}}, \bibinfo {author} {\bibfnamefont {R.~S.~K.}\ \bibnamefont
  {Mong}}, \bibinfo {author} {\bibfnamefont {C.}~\bibnamefont {Karrasch}},
  \bibinfo {author} {\bibfnamefont {J.~E.}\ \bibnamefont {Moore}}, \ and\
  \bibinfo {author} {\bibfnamefont {F.}~\bibnamefont {Pollmann}},\ }\href
  {\doibase 10.1103/PhysRevB.91.165112} {\bibfield  {journal} {\bibinfo
  {journal} {Phys. Rev. B}\ }\textbf {\bibinfo {volume} {91}},\ \bibinfo
  {pages} {165112} (\bibinfo {year} {2015})}\BibitemShut {NoStop}%
\bibitem [{ite()}]{itensor}%
  \BibitemOpen
  \href@noop {} {}\bibinfo {note} {{http://itensor.org}}\BibitemShut {NoStop}%
\bibitem [{\citenamefont {Rojo}\ \emph {et~al.}(1990)\citenamefont {Rojo},
  \citenamefont {Sofo},\ and\ \citenamefont {Balseiro}}]{RSB90}%
  \BibitemOpen
  \bibfield  {author} {\bibinfo {author} {\bibfnamefont {A.~G.}\ \bibnamefont
  {Rojo}}, \bibinfo {author} {\bibfnamefont {J.~O.}\ \bibnamefont {Sofo}}, \
  and\ \bibinfo {author} {\bibfnamefont {C.~A.}\ \bibnamefont {Balseiro}},\
  }\href {\doibase 10.1103/PhysRevB.42.10241} {\bibfield  {journal} {\bibinfo
  {journal} {Phys. Rev. B}\ }\textbf {\bibinfo {volume} {42}},\ \bibinfo
  {pages} {10241} (\bibinfo {year} {1990})}\BibitemShut {NoStop}%
\bibitem [{\citenamefont {Mentink}\ \emph {et~al.}(2015)\citenamefont
  {Mentink}, \citenamefont {Balzer},\ and\ \citenamefont {Eckstein}}]{MBE15}%
  \BibitemOpen
  \bibfield  {author} {\bibinfo {author} {\bibfnamefont {J.}~\bibnamefont
  {Mentink}}, \bibinfo {author} {\bibfnamefont {K.}~\bibnamefont {Balzer}}, \
  and\ \bibinfo {author} {\bibfnamefont {M.}~\bibnamefont {Eckstein}},\ }\href
  {https://doi.org/10.1038/ncomms7708} {\bibfield  {journal} {\bibinfo
  {journal} {Nat. Commun.}\ }\textbf {\bibinfo {volume} {6}},\ \bibinfo {pages}
  {6708} (\bibinfo {year} {2015})}\BibitemShut {NoStop}%
\bibitem [{\citenamefont {Peierls}(1933)}]{Pe33}%
  \BibitemOpen
  \bibfield  {author} {\bibinfo {author} {\bibfnamefont {R.}~\bibnamefont
  {Peierls}},\ }\href {\doibase 10.1007/BF01342591} {\bibfield  {journal}
  {\bibinfo  {journal} {Z. Phys.}\ }\textbf {\bibinfo {volume} {80}},\ \bibinfo
  {pages} {763} (\bibinfo {year} {1933})}\BibitemShut {NoStop}%
\bibitem [{\citenamefont {Takahashi}\ \emph {et~al.}(2008)\citenamefont
  {Takahashi}, \citenamefont {Itoh},\ and\ \citenamefont {Aihara}}]{TIA08}%
  \BibitemOpen
  \bibfield  {author} {\bibinfo {author} {\bibfnamefont {A.}~\bibnamefont
  {Takahashi}}, \bibinfo {author} {\bibfnamefont {H.}~\bibnamefont {Itoh}}, \
  and\ \bibinfo {author} {\bibfnamefont {M.}~\bibnamefont {Aihara}},\ }\href
  {\doibase 10.1103/PhysRevB.77.205105} {\bibfield  {journal} {\bibinfo
  {journal} {Phys. Rev. B}\ }\textbf {\bibinfo {volume} {77}},\ \bibinfo
  {pages} {205105} (\bibinfo {year} {2008})}\BibitemShut {NoStop}%
\bibitem [{\citenamefont {De~Filippis}\ \emph {et~al.}(2012)\citenamefont
  {De~Filippis}, \citenamefont {Cataudella}, \citenamefont {Nowadnick},
  \citenamefont {Devereaux}, \citenamefont {Mishchenko},\ and\ \citenamefont
  {Nagaosa}}]{FCNetal12}%
  \BibitemOpen
  \bibfield  {author} {\bibinfo {author} {\bibfnamefont {G.}~\bibnamefont
  {De~Filippis}}, \bibinfo {author} {\bibfnamefont {V.}~\bibnamefont
  {Cataudella}}, \bibinfo {author} {\bibfnamefont {E.~A.}\ \bibnamefont
  {Nowadnick}}, \bibinfo {author} {\bibfnamefont {T.~P.}\ \bibnamefont
  {Devereaux}}, \bibinfo {author} {\bibfnamefont {A.~S.}\ \bibnamefont
  {Mishchenko}}, \ and\ \bibinfo {author} {\bibfnamefont {N.}~\bibnamefont
  {Nagaosa}},\ }\href {\doibase 10.1103/PhysRevLett.109.176402} {\bibfield
  {journal} {\bibinfo  {journal} {Phys. Rev. Lett.}\ }\textbf {\bibinfo
  {volume} {109}},\ \bibinfo {pages} {176402} (\bibinfo {year}
  {2012})}\BibitemShut {NoStop}%
\bibitem [{\citenamefont {Lu}\ \emph {et~al.}(2012)\citenamefont {Lu},
  \citenamefont {Sota}, \citenamefont {Matsueda}, \citenamefont {Bon\v{c}a},\
  and\ \citenamefont {Tohyama}}]{LSMetal12}%
  \BibitemOpen
  \bibfield  {author} {\bibinfo {author} {\bibfnamefont {H.}~\bibnamefont
  {Lu}}, \bibinfo {author} {\bibfnamefont {S.}~\bibnamefont {Sota}}, \bibinfo
  {author} {\bibfnamefont {H.}~\bibnamefont {Matsueda}}, \bibinfo {author}
  {\bibfnamefont {J.}~\bibnamefont {Bon\v{c}a}}, \ and\ \bibinfo {author}
  {\bibfnamefont {T.}~\bibnamefont {Tohyama}},\ }\href {\doibase
  10.1103/PhysRevLett.109.197401} {\bibfield  {journal} {\bibinfo  {journal}
  {Phys. Rev. Lett.}\ }\textbf {\bibinfo {volume} {109}},\ \bibinfo {pages}
  {197401} (\bibinfo {year} {2012})}\BibitemShut {NoStop}%
\bibitem [{\citenamefont {Hashimoto}\ and\ \citenamefont
  {Ishihara}(2016)}]{HI16}%
  \BibitemOpen
  \bibfield  {author} {\bibinfo {author} {\bibfnamefont {H.}~\bibnamefont
  {Hashimoto}}\ and\ \bibinfo {author} {\bibfnamefont {S.}~\bibnamefont
  {Ishihara}},\ }\href {\doibase 10.1103/PhysRevB.93.165133} {\bibfield
  {journal} {\bibinfo  {journal} {Phys. Rev. B}\ }\textbf {\bibinfo {volume}
  {93}},\ \bibinfo {pages} {165133} (\bibinfo {year} {2016})}\BibitemShut
  {NoStop}%
\bibitem [{\citenamefont {Wang}\ \emph {et~al.}(2017)\citenamefont {Wang},
  \citenamefont {Claassen}, \citenamefont {Moritz},\ and\ \citenamefont
  {Devereaux}}]{WCMetal17}%
  \BibitemOpen
  \bibfield  {author} {\bibinfo {author} {\bibfnamefont {Y.}~\bibnamefont
  {Wang}}, \bibinfo {author} {\bibfnamefont {M.}~\bibnamefont {Claassen}},
  \bibinfo {author} {\bibfnamefont {B.}~\bibnamefont {Moritz}}, \ and\ \bibinfo
  {author} {\bibfnamefont {T.~P.}\ \bibnamefont {Devereaux}},\ }\href {\doibase
  10.1103/PhysRevB.96.235142} {\bibfield  {journal} {\bibinfo  {journal} {Phys.
  Rev. B}\ }\textbf {\bibinfo {volume} {96}},\ \bibinfo {pages} {235142}
  (\bibinfo {year} {2017})}\BibitemShut {NoStop}%
\bibitem [{\citenamefont {Eckstein}\ and\ \citenamefont {Werner}(2011)}]{EW11}%
  \BibitemOpen
  \bibfield  {author} {\bibinfo {author} {\bibfnamefont {M.}~\bibnamefont
  {Eckstein}}\ and\ \bibinfo {author} {\bibfnamefont {P.}~\bibnamefont
  {Werner}},\ }\href {\doibase 10.1103/PhysRevB.84.035122} {\bibfield
  {journal} {\bibinfo  {journal} {Phys. Rev. B}\ }\textbf {\bibinfo {volume}
  {84}},\ \bibinfo {pages} {035122} (\bibinfo {year} {2011})}\BibitemShut
  {NoStop}%
\bibitem [{\citenamefont {Werner}\ \emph {et~al.}(2014)\citenamefont {Werner},
  \citenamefont {Held},\ and\ \citenamefont {Eckstein}}]{WHE14}%
  \BibitemOpen
  \bibfield  {author} {\bibinfo {author} {\bibfnamefont {P.}~\bibnamefont
  {Werner}}, \bibinfo {author} {\bibfnamefont {K.}~\bibnamefont {Held}}, \ and\
  \bibinfo {author} {\bibfnamefont {M.}~\bibnamefont {Eckstein}},\ }\href
  {\doibase 10.1103/PhysRevB.90.235102} {\bibfield  {journal} {\bibinfo
  {journal} {Phys. Rev. B}\ }\textbf {\bibinfo {volume} {90}},\ \bibinfo
  {pages} {235102} (\bibinfo {year} {2014})}\BibitemShut {NoStop}%
\bibitem [{\citenamefont {Yanagiya}\ \emph {et~al.}(2015)\citenamefont
  {Yanagiya}, \citenamefont {Tanaka},\ and\ \citenamefont {Yonemitsu}}]{YTY15}%
  \BibitemOpen
  \bibfield  {author} {\bibinfo {author} {\bibfnamefont {H.}~\bibnamefont
  {Yanagiya}}, \bibinfo {author} {\bibfnamefont {Y.}~\bibnamefont {Tanaka}}, \
  and\ \bibinfo {author} {\bibfnamefont {K.}~\bibnamefont {Yonemitsu}},\ }\href
  {\doibase 10.7566/JPSJ.84.094705} {\bibfield  {journal} {\bibinfo  {journal}
  {J. Phys. Soc. Jpn.}\ }\textbf {\bibinfo {volume} {84}},\ \bibinfo {pages}
  {094705} (\bibinfo {year} {2015})}\BibitemShut {NoStop}%
\bibitem [{\citenamefont {Yang}(1989)}]{Ya89}%
  \BibitemOpen
  \bibfield  {author} {\bibinfo {author} {\bibfnamefont {C.~N.}\ \bibnamefont
  {Yang}},\ }\href {\doibase 10.1103/PhysRevLett.63.2144} {\bibfield  {journal}
  {\bibinfo  {journal} {Phys. Rev. Lett.}\ }\textbf {\bibinfo {volume} {63}},\
  \bibinfo {pages} {2144} (\bibinfo {year} {1989})}\BibitemShut {NoStop}%
\bibitem [{\citenamefont {Essler}\ \emph {et~al.}(1991)\citenamefont {Essler},
  \citenamefont {Korepin},\ and\ \citenamefont {Schoutens}}]{EKS91}%
  \BibitemOpen
  \bibfield  {author} {\bibinfo {author} {\bibfnamefont {F.~H.~L.}\
  \bibnamefont {Essler}}, \bibinfo {author} {\bibfnamefont {V.~E.}\
  \bibnamefont {Korepin}}, \ and\ \bibinfo {author} {\bibfnamefont
  {K.}~\bibnamefont {Schoutens}},\ }\href {\doibase
  10.1103/PhysRevLett.67.3848} {\bibfield  {journal} {\bibinfo  {journal}
  {Phys. Rev. Lett.}\ }\textbf {\bibinfo {volume} {67}},\ \bibinfo {pages}
  {3848} (\bibinfo {year} {1991})}\BibitemShut {NoStop}%
\bibitem [{\citenamefont {Essler}\ \emph {et~al.}(1992)\citenamefont {Essler},
  \citenamefont {Korepin},\ and\ \citenamefont {Schoutens}}]{EKS92}%
  \BibitemOpen
  \bibfield  {author} {\bibinfo {author} {\bibfnamefont {F.~H.}\ \bibnamefont
  {Essler}}, \bibinfo {author} {\bibfnamefont {V.~E.}\ \bibnamefont {Korepin}},
  \ and\ \bibinfo {author} {\bibfnamefont {K.}~\bibnamefont {Schoutens}},\
  }\href {\doibase 10.1016/0550-3213(92)90366-J} {\bibfield  {journal}
  {\bibinfo  {journal} {Nucl. Phys. B}\ }\textbf {\bibinfo {volume} {372}},\
  \bibinfo {pages} {559} (\bibinfo {year} {1992})}\BibitemShut {NoStop}%
\bibitem [{\citenamefont {Essler}\ \emph {et~al.}(2005)\citenamefont {Essler},
  \citenamefont {Frahm}, \citenamefont {G{\"o}hmann}, \citenamefont
  {Kl{\"u}mper},\ and\ \citenamefont {Korepin}}]{EFGetal05}%
  \BibitemOpen
  \bibfield  {author} {\bibinfo {author} {\bibfnamefont {F.~H.}\ \bibnamefont
  {Essler}}, \bibinfo {author} {\bibfnamefont {H.}~\bibnamefont {Frahm}},
  \bibinfo {author} {\bibfnamefont {F.}~\bibnamefont {G{\"o}hmann}}, \bibinfo
  {author} {\bibfnamefont {A.}~\bibnamefont {Kl{\"u}mper}}, \ and\ \bibinfo
  {author} {\bibfnamefont {V.~E.}\ \bibnamefont {Korepin}},\ }\href@noop {}
  {\emph {\bibinfo {title} {The One-Dimensional Hubbard Model}}}\ (\bibinfo
  {publisher} {Cambridge University Press},\ \bibinfo {address} {Cambridge,
  England},\ \bibinfo {year} {2005})\BibitemShut {NoStop}%
\bibitem [{NL()}]{NL}%
  \BibitemOpen
  \href@noop {} {}\bibinfo {note} {$\mathcal{C}_{N_{\eta}} = N_{\eta} !
  \prod^{N_{\eta}}_{l=1} (L-2N_0-l+1)$~\cite{Ta99}}\BibitemShut {NoStop}%
\bibitem [{\citenamefont {Takahashi}(1999)}]{Ta99}%
  \BibitemOpen
  \bibfield  {author} {\bibinfo {author} {\bibfnamefont {M.}~\bibnamefont
  {Takahashi}},\ }\href@noop {} {\emph {\bibinfo {title} {Thermodynamics of One
  Dimensional Solvable Models}}}\ (\bibinfo  {publisher} {Cambridge University
  Press},\ \bibinfo {address} {Cambridge, England},\ \bibinfo {year}
  {1999})\BibitemShut {NoStop}%
\bibitem [{EEE()}]{EEE}%
  \BibitemOpen
  \href@noop {} {}\bibinfo {note} {Note that $\ket{\psi _{N_{\eta }}}$ is the
  exact eigenstate of $\hat{\mathcal{H}}$ with the eigenenergy $E_{N_0} +
  N_{\eta } U$, where $E_{N_0}$ is the eigenenergy of $\ket{\psi^{({\rm
  GS})}_{N_0,N_0}}$.}\BibitemShut {Stop}%
\bibitem [{\citenamefont {Sakurai}(1994)}]{JJSakurai}%
  \BibitemOpen
  \bibfield  {author} {\bibinfo {author} {\bibfnamefont {J.~J.}\ \bibnamefont
  {Sakurai}},\ }\href@noop {} {\emph {\bibinfo {title} {Modern Quantum
  Mechanics}}}\ (\bibinfo  {publisher} {Addison-Wesley},\ \bibinfo {address}
  {Reading, MA},\ \bibinfo {year} {1994})\BibitemShut {NoStop}%
\bibitem [{\citenamefont {Rose}(1967)}]{MERose}%
  \BibitemOpen
  \bibfield  {author} {\bibinfo {author} {\bibfnamefont {M.~E.}\ \bibnamefont
  {Rose}},\ }\href@noop {} {\emph {\bibinfo {title} {Elementary Theory of
  Angular Momentum}}}\ (\bibinfo  {publisher} {Wiley},\ \bibinfo {address} {New
  York},\ \bibinfo {year} {1967})\BibitemShut {NoStop}%
\bibitem [{\citenamefont {Mazurenko}\ \emph {et~al.}(2017)\citenamefont
  {Mazurenko}, \citenamefont {Chiu}, \citenamefont {Ji}, \citenamefont
  {Parsons}, \citenamefont {Kanasz-Nagy}, \citenamefont {Schmidt},
  \citenamefont {Grusdt}, \citenamefont {Demler}, \citenamefont {Greif},\ and\
  \citenamefont {Greiner}}]{MCJetal17}%
  \BibitemOpen
  \bibfield  {author} {\bibinfo {author} {\bibfnamefont {A.}~\bibnamefont
  {Mazurenko}}, \bibinfo {author} {\bibfnamefont {C.~S.}\ \bibnamefont {Chiu}},
  \bibinfo {author} {\bibfnamefont {G.}~\bibnamefont {Ji}}, \bibinfo {author}
  {\bibfnamefont {M.~F.}\ \bibnamefont {Parsons}}, \bibinfo {author}
  {\bibfnamefont {M.}~\bibnamefont {Kanasz-Nagy}}, \bibinfo {author}
  {\bibfnamefont {R.}~\bibnamefont {Schmidt}}, \bibinfo {author} {\bibfnamefont
  {F.}~\bibnamefont {Grusdt}}, \bibinfo {author} {\bibfnamefont
  {E.}~\bibnamefont {Demler}}, \bibinfo {author} {\bibfnamefont
  {D.}~\bibnamefont {Greif}}, \ and\ \bibinfo {author} {\bibfnamefont
  {M.}~\bibnamefont {Greiner}},\ }\href {\doibase 10.1038/nature22362}
  {\bibfield  {journal} {\bibinfo  {journal} {Nature (London)}\ }\textbf
  {\bibinfo {volume} {545}},\ \bibinfo {pages} {462} (\bibinfo {year}
  {2017})}\BibitemShut {NoStop}%
\end{thebibliography}%

\clearpage

\appendix
\renewcommand\thesection{}
\renewcommand{\theequation}{S\arabic{equation}}
\setcounter{equation}{0}
\renewcommand\thefigure{S.\arabic{figure}}
\setcounter{figure}{0}
\renewcommand{\bibnumfmt}[1]{[S#1]}
\renewcommand{\citenumfont}[1]{S#1}

\section*{Supplemental Material}


\subsection{Exact diagonalization method}

To evaluate the state $\ket{\Psi(t)}$ under the time-dependent Hamiltonian $\hat{\mathcal{H}}(t)$, 
we numerically solve the time-dependent Schr\"odinger equation, 
\begin{equation}
i \frac{\partial}{\partial t}\ket{\Psi(t)} = \hat{\mathcal{H}}(t) \ket{\Psi(t)},
\end{equation}
with the initial condition that $\ket{\Psi(t=0)} = \ket{\psi_0}$, where 
$\ket{\psi_0}$ is the ground state of the Hamiltonian 
$\hat{\cal{H}}(t=0)$. For this purpose, 
we employ the time-dependent exact diagonalization (ED) method based on the Lanczos algorithm~\citeSM{PL86S,MA06S}.
In this method, the time evolution with a short time step $\delta t$ is calculated as 
\begin{eqnarray}
\ket{\Psi(t+\delta t)} &\simeq& e^{-i \hat{\mathcal{H}}(t) \delta t} \ket{\Psi(t)} \nonumber \\
&\simeq& \sum_{\ell=1}^{M_{\rm L}} e^{-i\xi_{\ell} \delta t} \ket{\tilde{\psi}_{\ell}} \braket{\tilde{\psi}_{\ell} |\Psi(t)}, 
\end{eqnarray}
where $\xi_{\ell}$ and $\ket{\tilde{\psi}_{\ell}}$ are eigenenergies and eigenvectors of $\hat{\mathcal{H}}(t)$, respectively, in the corresponding Krylov 
subspace generated with $M_{\rm L}$ Lanczos iterations~\citeSM{HI16S,PL86S,MA06S}. 
In our ED calculations, 
we adopt $\delta t = 0.01 / t_h$ and $M_{\rm L} = 15$ for the time evolution, which provides results with almost machine precision accuracy.


\subsection{One-dimensional (1D) Hubbard model with larger $L$: a MPS study}

\subsubsection*{Method}

In order to confirm the enhancement of the pair correlation in 
larger systems, we also perform the time-dependent matrix-product state (MPS)~\citeSM{PVWetal07S}
simulation for the time evolution starting from the ground state of the Hubbard model 
$\hat{\cal{H}}$ calculated by the density-matrix 
renormalization group method~\citeSM{Wh92S, Sc11S}. 
For the time evolution simulation, we employ the method proposed 
in Ref.~\citeSM{ZMKetal15S}, in which the time evolution operator is factorized 
as a compact form of the matrix product operator (MPO) representation.
In this method, the higher order approximation for the time evolution operator 
with time step $\delta t$ are formulated by introducing the additional set of 
time steps $\{ \delta t_1, \delta t_2, \cdots, \delta t_n \}$ with complex numbers 
in order to eliminate the unnecessary lowest order terms arisen from 
the MPO factorization. The resulting error is $\mathcal{O}(L\delta t^{p})$, 
where $L$ and $p$ denote the system size and the order of the approximation, 
respectively. Our calculation sets $p=3$, which requires the additional $n=4$ time steps, i.e., 
$\delta t_1= a + i b$, $\delta t_2 = a - i b$, $\delta t_3 = b+ i a$, 
and $\delta t_4 = b - i a$, with $a = (3+\sqrt{3})/12$ and $b = (3-\sqrt{3})/12$. 

\begin{figure}[tb]
\begin{center}
\includegraphics[width=0.65\columnwidth]{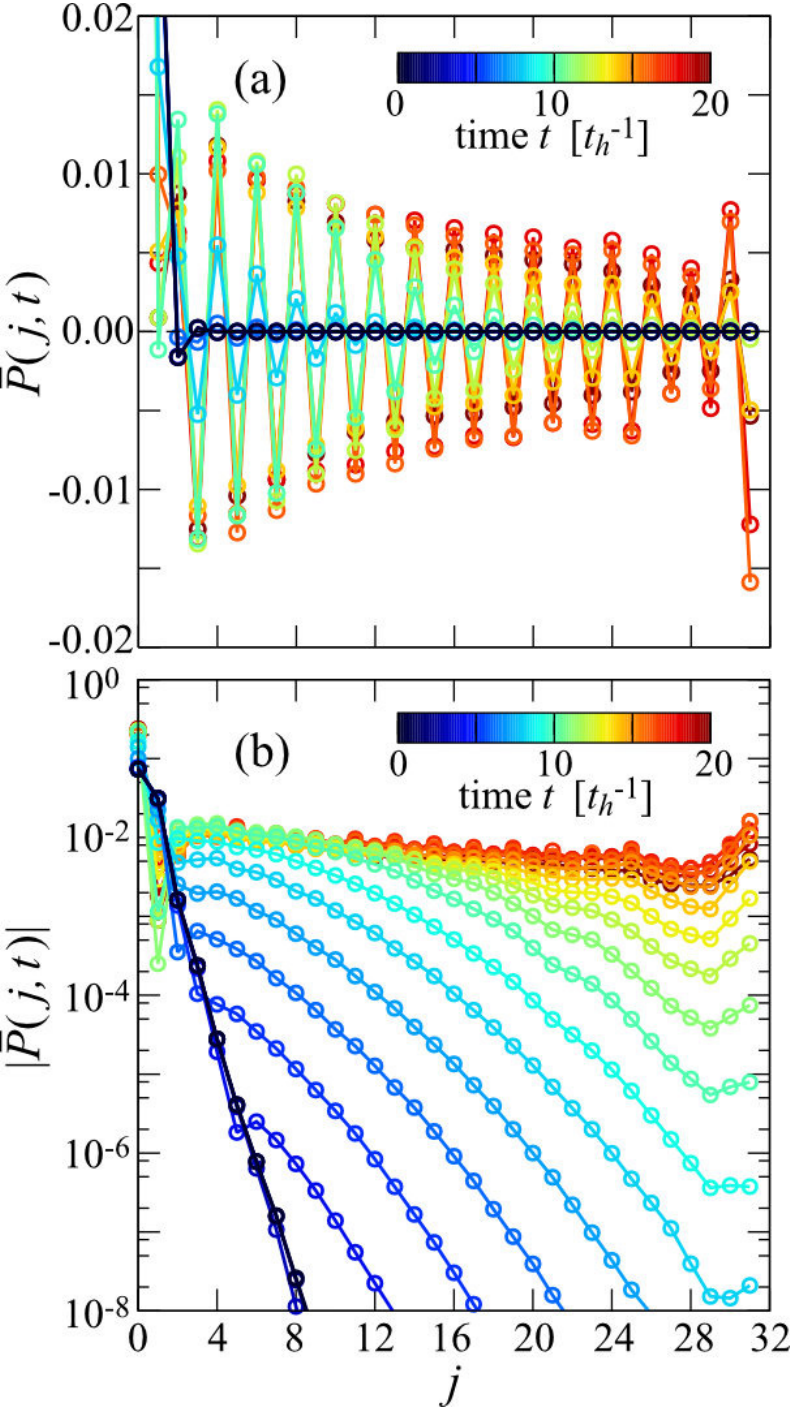}
\caption{
Time dependence of the on-site pair correlation function (a) $\bar{P}(j,t)$ and (b) logarithm of $|\bar{P}(j,t)|$ 
calculated by the time-dependent MPS method for a chain of $L=32$ sites with OBC at $U=8t_h$. 
Here, $A_0=0.2$, $\omega_p=8.26t_h$, $\sigma_p=2/t_h$, and $t_0=8/t_h$ are adopted in the vector potential $A(t)$. 
}
\label{s-fig1}
\end{center}
\end{figure}

For the MPS simulation, we use the ITensor package~\citeSM{itensorS}.
We keep the bond dimension up to $m=1200$ to calculate the ground state of $\hat{\cal{H}}$ for the initial state 
and $m=4800$ for the time evolution of the $L=32$ system under open boundary conditions (OBC).
The time step $\delta t$ is set to be $\delta t = 0.01/t_h$.

\subsubsection*{Results}

Figure~\ref{s-fig1} shows the real-space on-site pair correlation function 
\begin{equation}
\bar{P}(j,t) = \frac{1}{N_b}\sum^{N_b}_{i=1}  \bra{\Psi(t)} \left( \hat{\Delta}_{i+j}^\dag  
 \hat{\Delta}_{i} + {\rm H.c.} \right)  \ket{\Psi(t)}, 
\end{equation}
where $\hat{\Delta}_{i} = \hat{c}_{i,\uparrow} \hat{c}_{i,\downarrow}$ and $N_b = L-j$  
is the number of pairs of sites separated by distance $j$ in the system of $L$ sites with OBC. 
As shown in Fig.~\ref{s-fig1}, 
the pair correlation extends to a longer distance gradually with time in the transient period 
and shows clearly the sign-alternating feature that is characteristic of the $\eta$-pairing. 
The pair correlation eventually reaches to the longest distance in the system, 
similar to the results shown in Figs.~1(a) and 1(b) in the main text.


\subsection{$\eta$-pairing in the 1D Hubbard model for $L=14$} 

\begin{figure}[t]
\begin{center}
\includegraphics[width=\columnwidth]{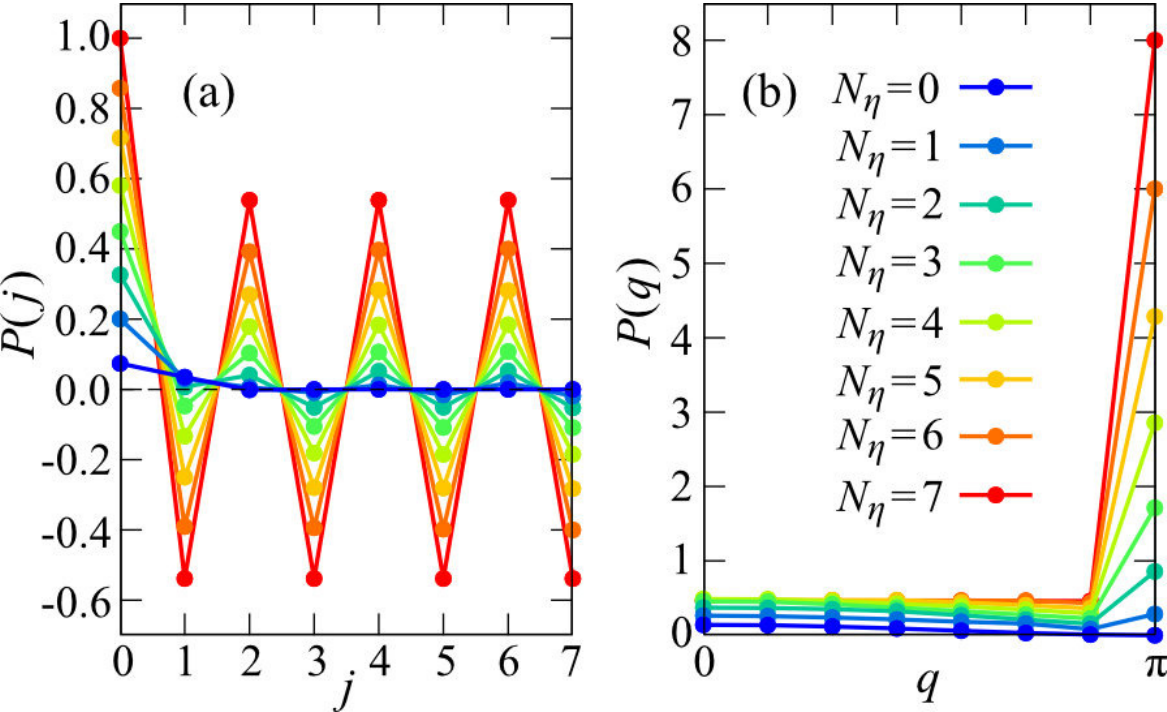}
\caption{
(a) On-site pair correlation function $P(j)$ and (b) on-site pair structure factor $P(q)$ 
for the half-filled $\eta$-paring eigenstate $\ket{\psi_{N_{\eta}}}$ 
at $U=8t_h$ with the different number of $\eta$ pairs $N_{\eta}$ ($\le L/2$).
$\ket{\psi_{N_{\eta}}}$ is generated from the ground state $\ket{\psi^{({\rm GS})}_{N_0,N_0}}$ of the Hubbard 
model $\hat{\cal{H}}$ with $N_0 = L/2-N_{\eta}$ calculated by the ED method for $L=14$ under PBC. 
}
\label{s-fig2}
\end{center}
\end{figure}

As an example, Fig.~3 in the main text shows the on-site pair correlation $P(j)$ and $P(q)$ 
of the $\eta$-pairing eigenstate 
\begin{equation}
\ket{\psi_{N_{\eta}}} = \frac{1}{\sqrt{\mathcal{C}_{N_{\eta}}}} (\hat{\eta}^{+})^{N_{\eta}} \ket{\psi^{({\rm GS})}_{N_0,N_0}}   
\end{equation} 
for $L=10$ simply because of the correspondence to Fig.~4(a) calculated for the 10 site cluster.
Here, we show supplementarily the results of $P(j)$ and $P(q)$ for $L=14$ at half-filling in Fig.~\ref{s-fig2}. 
The ground state 
$\ket{\psi^{({\rm GS})}_{N_0,N_0}}$ of the Hubbard model $\hat{\cal{H}}$ with 
$N_\uparrow = N_\downarrow = N_0 = L/2-N_{\eta}$ is calculated 
by the ED method under periodic boundary conditions (PBC). 
Note that $\ket{\psi_{N_{\eta}}}$ is the eigenstate of $\hat{\cal{H}}$ at half-filling with $N_\eta$ $\eta$ pairs. 
As shown in Fig.~\ref{s-fig2}, the density-wave-like pair correlation is largest for $N_{\eta}=L/2$. 

\subsection{Hubbard model on the square lattice}

In the main text, we focus on the 1D Hubbard model to demonstrate that the strong superconducting correlation 
can be induced in the Mott insulator (MI) by the pulse irradiation, and show that the origin of this 
superconductivity is 
due to the $\eta$-pairing mechanism. 
Here, we show that exactly the same conclusion can be reached for the two-dimensional (2D) Hubbard model on the 
square lattice with only nearest neighbor hoppings. 

\subsubsection*{Model and $\eta$ operators}\label{sec:model2d}

The 2D Hubbard model is described by the following Hamiltonian: 
\begin{equation}
{\hat {\mathcal{H}}} = - t_h \sum_{\langle i, j\rangle}\sum_\sigma  \left( {\hat c}_{i,\sigma}^{\dag}{\hat c}_{j,\sigma} + {\rm H.c.} \right)  
 + U \sum_{i}  {\hat n}_{i, \uparrow}  {\hat n}_{i, \downarrow}, 
 \label{eq:2d}
\end{equation}
where the sum $\langle i, j\rangle$ runs over all pairs of nearest neighbor sites $i$ and $j$ on the square lattice. 
Similarly to the 1D case, the total $\hat{\eta}$ operators $\hat{\eta}^\pm=\sum_j\hat{\eta}_j^\pm$ and 
$\hat{\eta}_z=\sum_j\hat{\eta}_j^z$ are defined in terms of 
the local operators 
$\hat{\eta}^+_j=(-1)^{j_x+j_y} \hat{c}^{\dag}_{j,\downarrow} \hat{c}^{\dag}_{j,\uparrow}$, 
$\hat{\eta}^-_j=(-1)^{j_x+j_y} \hat{c}_{j,\uparrow} \hat{c}_{j,\downarrow}, $ and 
$\hat{\eta}^z_j=\frac{1}{2}\left( {\hat n}_{j, \uparrow} + {\hat n}_{j, \downarrow} -1 \right)$, where  the location of 
site $j$ is given as $\bm{R}_j=j_x\bm{e}_x+j_y\bm{e}_y$ and $\bm{e}_{x(y)}$ is the unit vector along the $x(y)$ direction. 
These operators satisfy the $SU(2)$ commutation relations.  
We can also show that $[\hat{\mathcal{H}},\hat{\eta}^{\pm}] = \pm U \hat{\eta}^{\pm}$ and 
$[ \hat{\cal{H}}, \hat{\eta}^+\hat{\eta}^- ] = [   \hat{\cal{H}}, \hat{\eta}_z ]=0$. Therefore, 
any eigenstate of the Hubbard model $\hat{\cal{H}}$ can be chosen 
also to be an  eigenstate $|\eta, \eta_z\rangle $ of 
${\hat{\eta}}^2=\frac{1}{2}\left( \hat{\eta}^+\hat{\eta}^- + \hat{\eta}^-\hat{\eta}^+  \right) + \hat{\eta}_z^2$ 
and $\hat{\eta}_z$ with the eigenvalues $\eta(\eta+1)$ and $\eta_z$, respectively, where 
$|\eta, \eta_z\rangle $ can take $\eta=0,1,2,\cdots,L/2$ and $\eta_z=-\eta,-\eta+1, \cdots, \eta$, assuming that the number $N_{\uparrow}$ 
of up electrons and the number $N_\downarrow$ of down electrons are the same and $L$ (even) 
is the total number of sites. 
At half-filling with $N_{\uparrow} = N_\downarrow=L/2$, the eigenstates are characterized with $\eta=0,1,2,\cdots,L/2$ and $\eta_z=0$, 
and the ground state $\ket{\psi_0}$ of the Hubbard model $\hat{\cal{H}}$ is $\eta=\eta_z=0$.

The real-space on-site pair correlation function for the time-evolved state $\ket{\Psi(t)}$ is defined as 
\begin{equation}
P(\bm{R}_j,t) = \frac{1}{L}\sum_i  \bra{\Psi(t)} \left( \hat{\Delta}_{\bm{R}_i+\bm{R}_j}^\dag  
 \hat{\Delta}_{\bm{R}_i} + {\rm H.c.} \right)  \ket{\Psi(t)}, 
\end{equation}
where $\hat{\Delta}_{\bm{R}_i} = \hat{c}_{i,\uparrow} \hat{c}_{i,\downarrow}$ and the pair structure factor 
in the momentum space is given as 
\begin{equation}
P(\bm{q},t) =   \sum_j e^{i\bm{q}\cdot\bm{R}_j} P(\bm{R}_j,t).   
\end{equation} 
Noticing that $\hat{\Delta}_{\bm{R}_j} = (-1)^{j_x+j_y}\hat{\eta}^-_j$, $P(\bm{q},t)$ at $\bm{q}=\bm{\pi}=(\pi,\pi)$ is  
\begin{eqnarray}
P(\bm{q}=\bm{\pi},t) &=&  \frac{2}{L}  \bra{\Psi(t)}  \hat{\eta}^+ \hat{\eta}^-  \ket{\Psi(t)} \\
&=& \frac{2}{L} \bra{\Psi(t)} \left( \hat{\eta}^2 - \hat{\eta}_z^2  + \hat{\eta}_z \right)    \ket{\Psi(t)}. 
\end{eqnarray}
The pair structure factor $P(\bm{q}=\bm{\pi})$ for $\ket{\eta,\eta_z}$ is thus 
$2\left[ \eta(\eta+1) - \eta_z(\eta_z-1)  \right]/L$. 

Any eigenstate $|\eta, \eta_z\rangle $ can be constructed from the LWS $\ket{\eta,-\eta}$ by repeatedly 
applying $\hat{\eta}^+$ because 
\begin{equation}
\hat{\eta}^+ \ket{\eta,\eta_z} = \sqrt{\eta(\eta+1)-\eta_z(\eta_z+1)}\ket{\eta,\eta_z+1}. 
\end{equation}
Since $\hat{\eta}^-\ket{\eta,-\eta}=0$ by definition, 
the LWS contains no $\eta$ pairs and 
$P(\bm{q}=\bm{\pi})=0$. 
Each time that $\hat{\eta}^+$ is applied from the LWS, the number of $\eta$ pairs increases by one, and 
the maximum number of $\eta$ pairs is obtained when $\eta_z=0$ (i.e., half-filling) for a given $\eta$, where 
$\bra{\eta,\eta_z=0}\hat{\eta}^+\hat{\eta}^-\ket{\eta,\eta_z=0}=\eta(\eta+1)$ and the number of $\eta$ pairs is $\eta$. 

The time-dependent external field is introduced in Eq.~(\ref{eq:2d}) by 
$ t_h  \hat{c}_{i,\sigma}^{\dag} \hat{c}_{j,\sigma} \rightarrow 
t_h e^{-i\bm{A}(t)\cdot (\bm{R}_i-\bm{R}_j)} \hat{c}_{i,\sigma}^{\dag} \hat{c}_{j,\sigma}$  
with the time-dependent vector potential $\bm{A}(t) = A(t)(\bm{e}_x+\bm{e}_y)$ pointing along the diagonal direction 
and $A(t)$  given in the main text. 
The current operator $\hat{J}_\alpha^{(0)}$ along a direction $\alpha$  ($\alpha=x,y$) is defined as 
\begin{equation}
\hat{J}_\alpha^{(0)}=it_h\sum_{j,\sigma} \left(\hat{c}^{\dag}_{j+\bm{e}_\alpha,\sigma} \hat{c}_{j,\sigma} - \hat{c}^{\dag}_{j,\sigma} \hat{c}_{j+\bm{e}_\alpha,\sigma}\right),
\end{equation}
where $\hat{c}^{\dag}_{j+\bm{e}_\alpha,\sigma}$ is the creation operator of an electron at the site located at 
$\bm{R}_j+\bm{e}_\alpha$ with spin $\sigma$. 
We can now show that 
\begin{equation}
\left[ \hat{\eta}^\pm, \hat{J}^{(0)}_\alpha \right] = \sqrt{2} \hat{J}^{(\pm1)}_\alpha, \;\; 
\left[ \hat{\eta}_z, \hat{J}^{(0)}_\alpha \right] = 0, 
\end{equation} 
\begin{equation}
\left[ \hat{\eta}^\pm, \hat{J}^{(\mp1)}_\alpha \right] = \sqrt{2} \hat{J}^{(0)}_\alpha, \;\;
\left[ \hat{\eta}_z, \hat{J}^{(\pm1)}_\alpha \right] = \pm \hat{J}^{(\pm1)}_\alpha, 
\end{equation} 
where 
\begin{equation}
\hat{J}^{(+1)}_\alpha =\sqrt{2}it_h\sum_j(-1)^{j_x+j_y}\bigl(\hat{c}^\dag_{j+\bm{e}_\alpha,\uparrow} \hat{c}^\dag_{j,\downarrow}   +\hat{c}^\dag_{j,\uparrow} \hat{c}^\dag_{j+\bm{e}_\alpha,\downarrow} \bigr), 
\end{equation}
and
\begin{equation}
\hat{J}^{(-1)}_\alpha =\sqrt{2}it_h\sum_j(-1)^{j_x+j_y}\bigl(\hat{c}_{j+\bm{e}_\alpha,\downarrow} \hat{c}_{j,\uparrow}   +\hat{c}_{j,\downarrow} \hat{c}_{j+\bm{e}_\alpha,\uparrow} \bigr).
\end{equation}
Therefore, $\hat{J}_\alpha^{(q)}$ with $q=0,\pm1$ is a rank-one tensor operator in terms of $\hat{\eta}$ operators.  
In particular, the current operator $\hat{J}^{(0)}_\alpha$ is a rank-one tensor operator with $q=0$ and hence 
there is the following selection rule: $\bra{\eta',\eta_z'}\hat{J}^{(0)}_\alpha\ket{\eta,\eta_z}\ne0$ only for 
$\eta'=\eta\pm1$ when $\eta_z'=\eta_z=0$~\citeSM{JJSakuraiS,MERoseS}. 
We also note that 
$it_h\sum_{\langle i,j\rangle}\sum_\sigma \sin \left[ \bm{A}(t)\cdot (\bm{R}_i-\bm{R}_j)  \right] 
\left( \hat{c}_{i,\sigma}^{\dag} \hat{c}_{j,\sigma} -  {\rm H.c.} \right)$
is a rank-one tensor operator with $q=0$, while 
$-t_h\sum_{\langle i,j\rangle}\sum_\sigma \cos \left[ \bm{A}(t)\cdot (\bm{R}_i-\bm{R}_j)  \right] 
\left( \hat{c}_{i,\sigma}^{\dag} \hat{c}_{j,\sigma} +{\rm H.c.} \right)$ is a rank-zero tensor operator, i.e., 
a scalar operator. 

Although here we consider the 2D case, the extension to other spatial dimensions is straightforward. 

\begin{figure}[t]
\begin{center}
\includegraphics[width=\columnwidth]{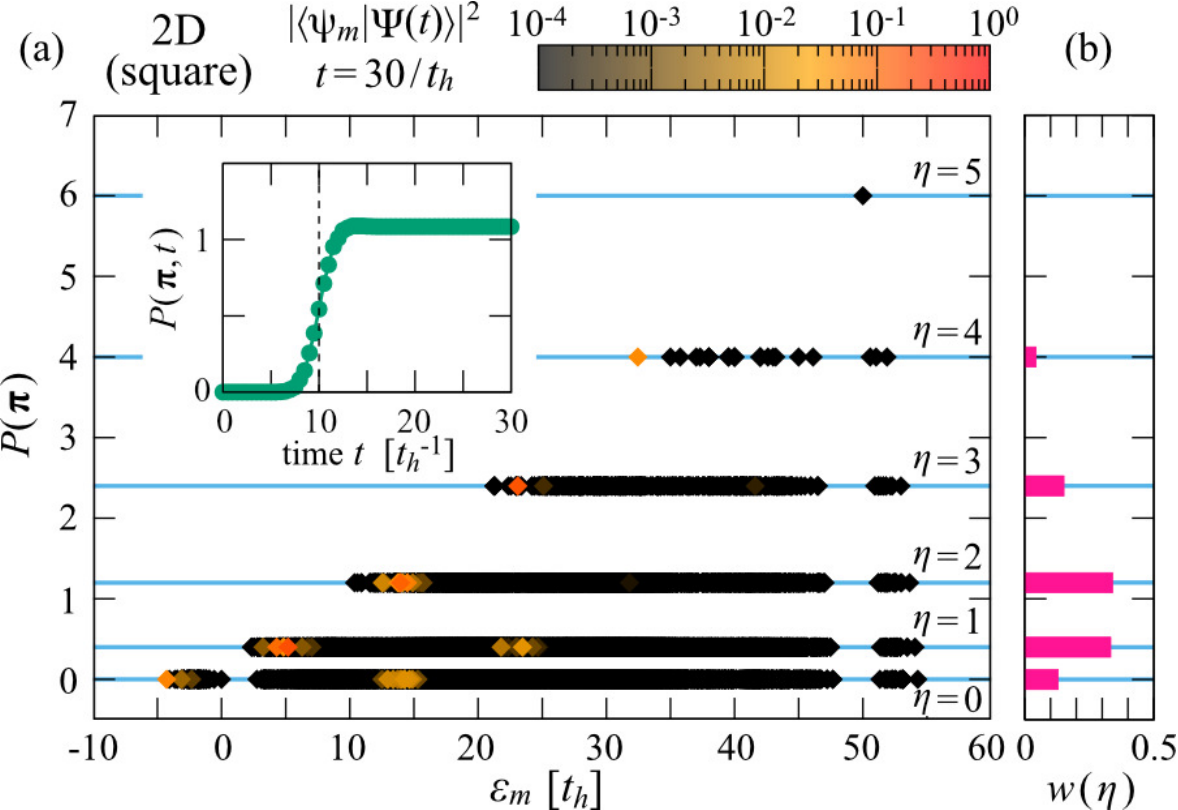}
\caption{
(a) All eigenenergies $\varepsilon_m$ and $P(\bm{q}=\bm{\pi})$ [$\bm{\pi}= (\pi,\pi)$] for the eigenstates 
$\ket{\psi_m}$ of the half-filled Hubbard model on a 
$\sqrt{10}\times\sqrt{10}$ cluster with PBC at $U=10t_h$.  
The color of each point (diamond) indicates the weight $|\braket{\psi_m|\Psi(t)}|^2$ of the eigenstate $\ket{\psi_m}$ 
in the photoinduced state $\ket{\Psi(t)}$ at $t=30/t_h$. Here, $A_0=0.25$, 
$\omega_p=9.1t_h$, $\sigma_p=2/t_h$, and  $t_0=10/t_h$ are adopted in the vector potential $A(t)$.
When the eigenstates are degenerate, the color indicates the sum of $|\braket{\psi_m|\Psi(t)}|^2$ over these degenerate 
states. 
The time evolution of $P(\bm{q}=\bm{\pi},t)$ for $\ket{\Psi(t)}$ is also shown in the inset. 
(b) The total weight $w(\eta)$ of $|\braket{\psi_m|\Psi(t)}|^2$ over the states $\ket{\psi_m}$ that have the same number 
$\eta$ of $\eta$ pairs, and thus $\sum_{\eta=0}^{L/2}w(\eta)=1$. The parameters are the same as in (a).  
}
\label{s-fig3}
\end{center}
\end{figure}

\subsubsection*{Results}

As shown above, 
any eigenstate of $\hat{\cal{H}}$ can be chosen to be 
an eigenstate of $\hat{\eta}^2$ and $\hat{\eta}_z$. 
Figure~\ref{s-fig3}(a) shows all the eigenenergies $\varepsilon_m$ of $\hat{\cal{H}}$ 
and the corresponding pair structure factors 
$P(\bm{q})$ at $\bm{q}=\bm{\pi}=(\pi,\pi)$ on a $\sqrt{10}\times \sqrt{10}$ cluster with PBC at half-filling. 
Indeed, as in the 1D case, $P(\bm{\pi})$ is quantized 
as $P(\bm{\pi}) = 2 \eta(\eta +1)/L$, where $\eta\,(=0,1,\cdots,L/2)$ corresponds to the number of $\eta$ pairs. 
As shown in Fig~\ref{s-fig3}(b), the photoinduced state $\ket{\Psi(t)}$ after the pulse irradiation displays 
nonzero overlaps with the eigenstates $\ket{\psi_m}$ of $\hat{\cal{H}}$ with $\eta\ne0$. 
This is responsible for the large enhancement of $P(\bm{\pi},t)$ in the photoinduced state $\ket{\Psi(t)}$ 
[see the inset of Fig.~\ref{s-fig3}(a)]. 
Since the current operator is a rank-one tensor operator, we can again observe in Fig.~\ref{s-fig3}(a) 
a ``tower of states" structure of the eigenstates $\ket{\psi_m}$ contributing
to the photoinduced state $\ket{\Psi(t)}$ with  large weights 
$|\braket{\psi_m|\Psi(t)}|^2$.


\subsection{1D Hubbard model away from half-filling} 

\begin{figure}[t]
\begin{center}
\includegraphics[width=\columnwidth]{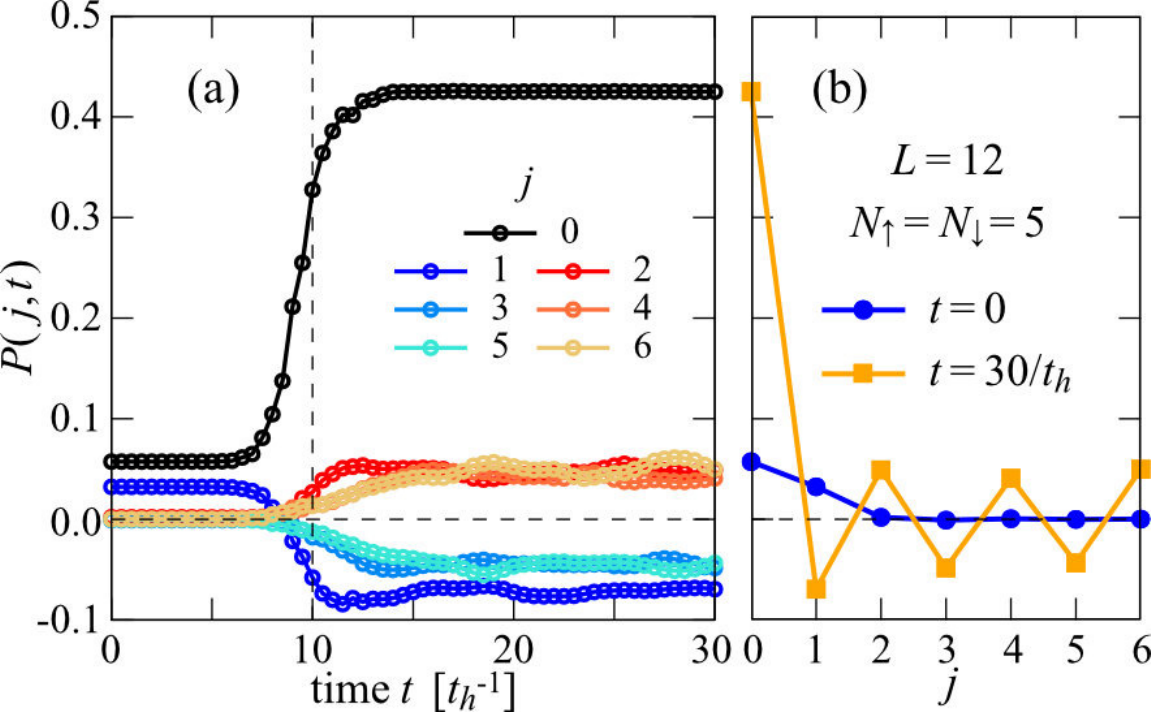}
\caption{
(a)~Time evolution of the on-site pair correlation function $P(j,t)$ with hole doping.  
(b)~$P(j,t)$ at $t=0$ (blue circles) and $t=30/t_h$ (orange squares). 
The results are calculated by the ED method for $L=12$ and $N_{\uparrow}=N_{\downarrow}=5$ at $U=8t_h$ with 
$\sigma_p=2/t_h$, $t_0=10/t_h$, $A_0=0.7$, and $\omega_p=8.8t_h$. 
}
\label{s-fig4}
\end{center}
\end{figure}

We also examine the behavior of the photoinduced states in the 1D Hubbard model $\hat{\cal{H}}$ away from half-filling. 
Figure~\ref{s-fig4} shows the time evolution of the pair correlation function $P(j,t)$ calculated by the ED method 
for $L=12$ with $N_{\uparrow}=N_{\downarrow}=5$ (10 electrons in total) under PBC. 
Although the magnitude of $P(j,t)$ is smaller than that for the case of half-filling, $P(j,t)$ clearly shows a pair density wave 
like oscillation with the correlation extended up to the longest distance of the cluster. 
Therefore, the $\eta$-pairing correlation is induced in the photoexcited state in the Hubbard model 
even away from half-filling. 

\begin{figure}[t]
\begin{center}
\includegraphics[width=\columnwidth]{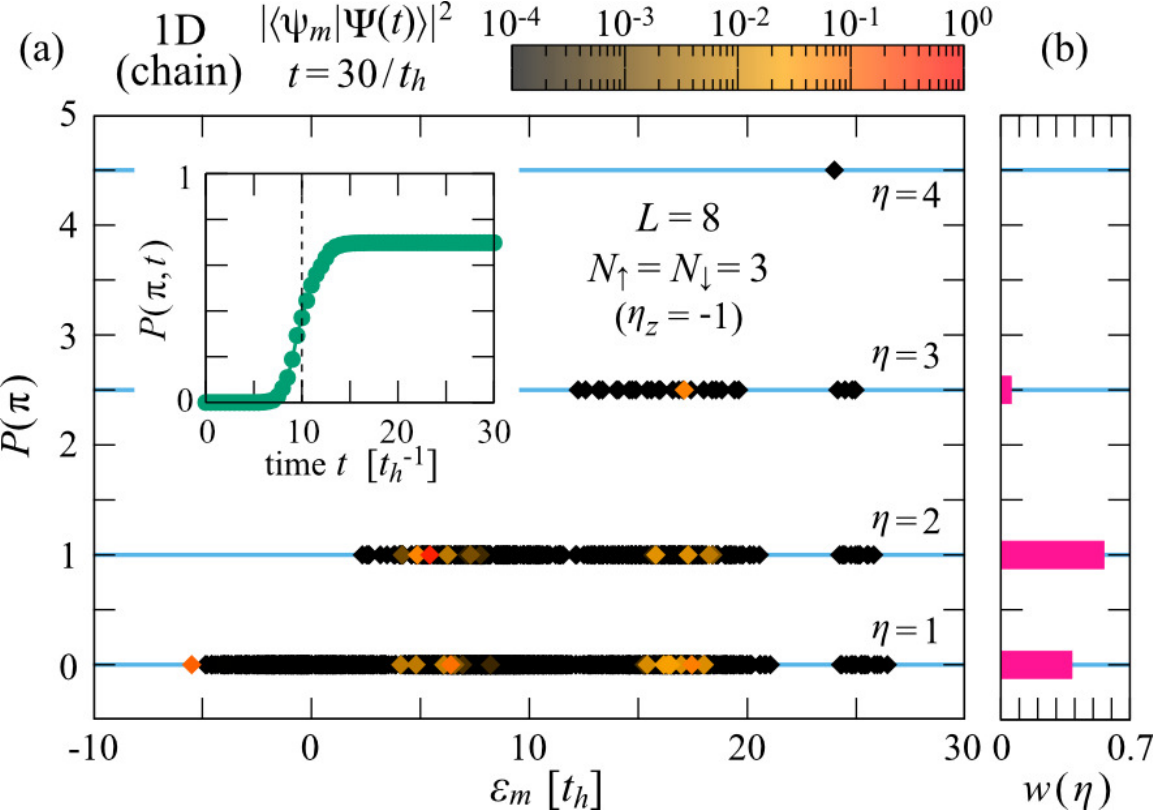}
\caption{
(a) All eigenenergies $\varepsilon_m$ and $P(q=\pi)$ for the eigenstates $\ket{\psi_m}$ of 
the hole-doped 1D Hubbard model $\hat{\cal{H}}$ at $U=8t_h$ for $L=8$ under PBC with 
$N_{\uparrow}=N_{\downarrow}=3$ electrons.
The color of each point (diamond) indicates the weight $|\braket{\psi_m|\Psi(t)}|^2$ of the eigenstate $\ket{\psi_m}$ 
in the photoinduced state $\ket{\Psi(t)}$ at $t=30/t_h$. Here, $A_0=0.7$, 
$\omega_p=10.7t_h$, $\sigma_p=2/t_h$, and  $t_0=10/t_h$ are adopted in the vector potential $A(t)$.
When the eigenstates are degenerate, the color indicates the sum of $|\braket{\psi_m|\Psi(t)}|^2$ over these degenerate 
states. 
The time evolution of $P(q=\pi,t)$ for $\ket{\Psi(t)}$ is also shown in the inset. 
(b) The total weight $w(\eta)$ of $|\braket{\psi_m|\Psi(t)}|^2$ over the states $\ket{\psi_m}$ that have the same 
eigenvalue of $\eta$,
and thus $\sum_{\eta=1}^{L/2}w(\eta)=1$.  
The parameters are the same as in (a).  
Note that the number of $\eta$ pairs is $\eta-1$ for this hole-doped case. 
}
\label{s-fig5}
\end{center}
\end{figure}

To elucidate the nature of the photoinduced state $\ket{\Psi(t)}$ in terms of the $\eta$ pairs, 
we calculate the eigenenergies $\varepsilon_m$ and the structure factors $P(q\!=\!\pi)$ 
for all the eigenstates $\ket{\psi_m}$ of the 1D Hubbard model $\hat{\cal{H}}$ with hole-doping. 
Figure~\ref{s-fig5} shows the results for $L=8$ with $N_{\uparrow}=N_{\downarrow}=3$ (6 electrons in total) under PBC. 
As shown in Fig.~\ref{s-fig5}(a), the structure factor $P(q\!=\!\pi)$ for each eigenstate is nicely quantized. 
This is because each eigenstate $\ket{\psi_m}$ away from half-filling is also the eigenstate of 
$\hat{\eta}^2$ and $\hat{\eta}_z$. 
The quantized values are given as
\begin{eqnarray} 
P(q\!=\!\pi) &=& \frac{2}{L} \! \bra{\psi_{m}}  \hat{\eta}^+\hat{\eta}^-  \ket{\psi_{m}} 
\!=\! \frac{2}{L} \! \bra{\psi_{m}} \! ( \hat{\eta}^2 \!-\! \hat{\eta}^2_z \!+\! \hat{\eta}_z ) \! \ket{\psi_{m}} 
\notag \\
&=& \frac{2}{L} [\eta(\eta +1)-\eta_z(\eta_z-1)]
\end{eqnarray}
with $\eta=|\eta_z|, |\eta_z|+1,\cdots,\frac{L}{2}$  
and $\eta_z = (N_{\uparrow}+N_{\downarrow}-L)/2=-1$. 
Note that $P(\pi)=0$ (no $\eta$ pair state) is characterized by the state with $\eta=1$ because $\eta_z=-1$ and this state  
is the LWS. 

In Fig.~\ref{s-fig5}(a), the color of each point indicates the weight $|\braket{\psi_m|\Psi(t)}|^2$ of the eigenstate $\ket{\psi_m}$ 
in the photoinduced state $\ket{\Psi(t)}$ that exhibits the enhancement of $P(q=\pi,t)$ after the pulse irradiation [see the inset of Fig~\ref{s-fig5}(a)].
We find that the state $\ket{\Psi(t)}$ after the pulse irradiation contains the nonzero weights of the eigenstates 
$\ket{\psi_m}$ with finite $P(\pi)$ [also see Fig.~\ref{s-fig5}(b)].  
Therefore, the reason for the enhancement of $P(q=\pi,t)$ is  
the same as in the case at half-filling. 

However, the distribution of the weight $|\braket{\psi_m|\Psi(t)}|^2$ after the pulse irradiation in Fig.~\ref{s-fig5}(a) is 
qualitatively different from that in the case at half-filling shown in Fig.~4(a) in the main text. 
For example, there is the finite contribution to the weight from the eigenstates 
with $P(\pi)=0$ around $\varepsilon_m-\varepsilon_0\sim \omega_p$, 
which is absent at half-filling. 
This is explained by the different selection rules of the current operator $\hat{J}$ for the half-filled ($\eta_z=0$) 
and hole-doped ($\eta_z \ne 0$) states. 
As mentioned in the main text and also in Sec.~\ref{sec:model2d}, 
$\hat{J}$ is a rank-one tensor operator with the zeroth component in terms of the $\hat{\eta}$ operators.  
Hence, from the Wigner--Eckart theorem~\citeSM{JJSakuraiS,MERoseS}, the selection rule of 
$\bra{\eta',\eta_z'}\hat{J}\ket{\eta,\eta_z}$ is given as 
\begin{equation}
\bra{\eta',\eta_z'}\hat{J}\ket{\eta,\eta_z} \propto
\left( \begin{array}{ccc}
\eta & 1 & \eta'           \\ 
\eta_z & 0 &  -\eta'_z  \\
\end{array} \right) 
\label{eq:3jrule}
\end{equation} 
with the 3$j$-symbol.
The 3$j$-symbol is zero unless $\eta-1 \le \eta' \le\eta+1$ and $\eta_z-\eta'_z=0$ are satisfied. 
Therefore, $\bra{\eta',\eta_z'}\hat{J}\ket{\eta,\eta_z}\ne0$ for $\eta'=\eta,\,\eta\pm1$ when $\eta'_z=\eta_z \ne 0$ 
for the hole-doped states. The result in Fig.~\ref{s-fig5}(a) follows this selection rule. 
However, when $\eta_z=\eta'_z=0$ for the half-filled state, the nonzero 3$j$-symbol must satisfy 
the additional rule: $\eta+\eta'+1 = ({\rm even})$. 
Therefore, the excitation to the states with $\eta'=\eta$ is not induced by $\hat{J}$ at half-filling 
($\eta_z=\eta'_z=0$), and $\bra{\eta',0}\hat{J}\ket{\eta,0}\ne0$ only for $\eta'=\eta\pm1$.
The results at half-filling in Fig.~4(a) in the main text and Fig.~\ref{s-fig3}(a) follow this selection rule.


\subsection{Perturbation analysis in the limit of large pulse width $\sigma_p$}

In the large pulse width limit, i.e., $\sigma_p \to \infty$, the time-dependent vector 
potential is given as $A(t)=A_0\cos\left[  \omega_p(t-t_0) \right]$. 
Let us denote the time-dependent Hamiltonian with the time-dependent external field as 
\begin{align}
\hat{\mathcal{H}}(t) = \hat{\mathcal{H}} + \hat{\mathcal{V}}(t),
\end{align}
where $\hat{\mathcal{H}}$ is the time-independent part of the Hamiltonian given by, e.g., 
Eq.~(1) in the main text 
and $\hat{\mathcal{V}}(t)$ is the time-dependent part of the Hamiltonian given as 
\begin{align}
\hat{\mathcal{V}}(t) = - t_h \sum_{j,\sigma}
\left( 
e^{iA(t)}-1 
\right) 
\hat{c}_{j,\sigma}^{\dagger} \hat{c}_{j+1,\sigma} + 
{\rm H.c.}. 
\end{align} 
Because $A(t)$ becomes a periodic function of $t$ in the limit $\sigma_p \to \infty$, 
$\hat{\mathcal{V}}(t)$ can be expanded using Bessel functions of the first kind $\mathcal{J}_{\mu}(x)$ 
($\mu$: integer)~\citeSM{KA16S}, i.e., 
\begin{align}
\hat{\mathcal{V}}(t) = \sum_{\mu=-\infty}^{\infty} \hat{\mathcal{V}}^{(\mu)} e^{-i \mu \omega_p t},
\label{eq:besselexpansion}
\end{align}
where
\begin{align}
\hat{\mathcal{V}}^{(0)} & = (\mathcal{J}_{0}(A_0)-1) \hat{K}, 
\notag \\
\hat{\mathcal{V}}^{(2\mu)} & = (-1)^{\mu} \mathcal{J}_{2\mu}(A_0) \hat{K}, \quad (\mu \neq 0) 
\label{eq:besselexpansion:terms} \\
\hat{\mathcal{V}}^{(2\mu+1)} & = (-1)^{\mu} \mathcal{J}_{2\mu+1}(A_0) \hat{J}. 
\notag 
\end{align}
Here we set  $t_0=0$. 
It is important to notice in Eqs.~(\ref{eq:besselexpansion}) and (\ref{eq:besselexpansion:terms}) 
that the operator $\hat{K}$ in the $\mu$ even terms is 
the kinetic (rank-zero tensor) operator, i.e.,  
\begin{align}
\hat{K} =  - t_h \sum_{j,\sigma}  \left( {\hat c}_{j,\sigma}^{\dag}{\hat c}_{j+1,\sigma} 
+ {\hat c}_{j+1,\sigma}^{\dag}{\hat c}_{j,\sigma} \right), 
\end{align}
while the operator $\hat{J}$ in the $\mu$ odd terms is 
the current (rank-one tensor) operator, i.e., 
\begin{align}
\hat{J} =  - i t_h \sum_{j,\sigma}  \left( {\hat c}_{j,\sigma}^{\dag}{\hat c}_{j+1,\sigma} - 
{\hat c}_{j+1,\sigma}^{\dag}{\hat c}_{j,\sigma}\right),  
\end{align}
as defined also in the main text. 
 
A time-dependent state $\vert \Psi (t) \rangle$ governed by $\hat{\cal{H}}(t)$ can be expanded as 
\begin{align}
\vert \Psi (t) \rangle = \sum_m c_m (t) \vert \psi_m \rangle ,
\end{align}
where $\vert \psi_m \rangle$ ($m=0,1,2,\cdots$) are 
the $m$th eigenstate of $\hat{\mathcal{H}}$ 
with the eigenenergy $\varepsilon_m$. For simplicity, 
we assume that the ground state is not degenerate with $\varepsilon_0 < \varepsilon_1 \leq \varepsilon_2 \leq \cdots$. 
By using the time-dependent perturbation theory, 
the coefficient $c_m(t)$ is obtained as the sum 
over terms $c_m^{(k)} (t)$ of the $k$th order expansion in terms of $\hat{\mathcal{V}}(t)$: 
\begin{align}
c_m (t) = \sum_{k=0}^{\infty} c_m^{(k)} (t). 
\end{align}
Assuming that the initial state at time $t_{\rm i} =-\infty$ 
is the ground state $\vert \psi_0 \rangle$ 
of $\hat{\mathcal{H}}$, $c_m^{(k)}(t)$ is given as  
\begin{widetext}
\begin{eqnarray}
c_m^{(k)}(t) = 
( - i )^k 
\int_{-\infty}^{t}  \! d t_k 
 \cdots \!
\int_{-\infty}^{t_{3}}  \! d t_2 
\int_{-\infty}^{t_{2}}  \! d t_1
\sum_{m_{k-1}}  
 \cdots \sum_{m_2}   \sum_{m_1}  
\braket{ \psi_m | \hat{\mathcal{V}}_I (t_k) | \psi_{m_{k-1}} } 
\cdots   \!
\braket{ \psi_{m_{2}} | \hat{\mathcal{V}}_I (t_2) | \psi_{m_{1}} }
\braket{ \psi_{m_{1}} | \hat{\mathcal{V}}_I (t_1) | \psi_{0} },
\notag \\
\end{eqnarray}
\end{widetext}
where $\hat{\mathcal{V}}_I(t) = e^{i\hat{\mathcal{H}}t} \hat{\mathcal{V}}(t) e^{- i\hat{\mathcal{H}}t}$~\citeSM{JJSakuraiS}.
Because of Eq.~(\ref{eq:besselexpansion}), 
\begin{align}
\braket{ \psi_m | \hat{\mathcal{V}}_I(t)| \psi_{m'} } 
&= \sum_{\mu}e^{i(\varepsilon_m-\varepsilon_{m'}-\mu\omega_p)t} \mathcal{V}^{(\mu)}_{mm'}
\label{HM_etaSR_eq94} 
\end{align}
with 
\begin{align}
\mathcal{V}_{mm^{\prime}}^{(\mu)} = \langle \psi_m \vert \hat{\mathcal{V}}^{(\mu)} \vert \psi_{m^{\prime}} \rangle.  
\end{align}
Therefore, we obtain for $t \to \infty$ that 
\begin{widetext}
\begin{eqnarray}
c_m^{(k)}(\infty) &=&  2 \pi i (-1)^k \sum_{\mu_k} \cdots \sum_{\mu_2} \sum_{\mu_1} 
\sum_{m_{k-1}} \cdots \sum_{m_2} \sum_{m_1} 
\mathcal{V}_{m m_{k-1}}^{(\mu_k)} \cdots 
\mathcal{V}_{m_2 m_1}^{(\mu_2)} 
\mathcal{V}_{m_1 0}^{(\mu_1)} 
\prod_{k^{\prime}=1}^{k-1} 
\frac{1}{\varepsilon_{m_{k^{\prime}}}-\varepsilon_0 - \left(\sum_{\ell=1}^{k^{\prime}} \mu_{\ell}\right) \omega_p - i \gamma}
\nonumber \\
& &\quad\quad \times \delta \left( \varepsilon_m - \varepsilon_0 - \left(\sum_{\ell=1}^k \mu_{\ell}\right) \omega_p \right),
\label{eq:perturbation}
\end{eqnarray}
\end{widetext}
where $\gamma \to 0^+$ is a convergence factor. 

It is now obvious from the delta function in Eq.~(\ref{eq:perturbation}) 
that the coefficients $c_m^{(k)}(t)$ for $t \to \infty$ 
can be nonzero only if $\varepsilon_m - \varepsilon_0 = \left(\sum_{\ell = 1}^k \mu_{\ell}\right) \omega_p$, 
suggesting that the excitations are allowed only to states with the excitation energy that is 
an integer multiple of $\omega_p$. 
This nicely explains the energy dependence found in Fig.~4(a) in the main text and Fig.~\ref{s-fig3}(a)  
for half-filling and also in Fig.~\ref{s-fig5}(a) away from half-filling. 
For example, if $\sum_{\ell} \mu_{\ell} = 2 \nu + 1$ ($\nu$: integer), 
$\mathcal{V}_{m m_{k-1}}^{(\mu_k)} \cdots 
\mathcal{V}_{m_{2} m_{1}}^{(\mu_2)} 
\mathcal{V}_{m_{1} 0}^{(\mu_1)}$ should involve the odd number of 
excitations induced by the current operator $\hat{J}$. 
In the case of half-filling, combining this with the selection rule in Eq.~(\ref{eq:3jrule}) 
yields that the $\eta$ odd excitations are 
possible if and only if $\varepsilon_m - \varepsilon_0 = (2 \nu + 1) \omega_p$. 
Similarly, the $\eta$ even excitations are possible if and only if 
the $\varepsilon_m - \varepsilon_0 = 2 \nu \omega_p$ at half-filling. 
These are in accordance with the ``tower of states'' structure shown schematically 
in Fig.~4(c) in the main text.


\subsection{1D Hubbard model with the next-nearest-neighbor hopping} 

In this and the next sections, we investigate the pair correlations when the $\eta$ commutation relations, e.g. 
$[\hat{\mathcal{H}},\hat{\eta}^{\pm}] = \pm U \hat{\eta}^{\pm}$, are broken  
in the Mott-Hubbard system. 
First, we consider the 1D Hubbard model with the next-nearest-neighbor (NNN) hopping $t'_h$ described by 
$\hat{\cal{H}}_{\rm NNN}=\hat{\cal{H}}+\hat{\mathcal{H}}_{t'_h}$, where $\cal{\hat{H}}$ is given by 
Eq.~(1) in the main text and 
\begin{equation}
\hat{\mathcal{H}}_{t'_h} = - t'_h \sum_{j,\sigma}  ( \hat{c}_{j,\sigma}^{\dag}\hat{c}_{j+2,\sigma} + {\rm H.c.} )
\end{equation}
is the NNN hopping term. Because  
$[\hat{\mathcal{H}}_{t'_h},\hat{\eta}^{+}] = -4t'_h \sum_{k} \cos(2k) \hat{c}^{\dag}_{\pi-k,\downarrow} \hat{c}^{\dag}_{k ,\uparrow} \ne 0$, the Hamiltonian $\hat{\mathcal{H}}_{\rm NNN}$ breaks the $\eta$ commutation relations. 

\begin{figure}[t]
\begin{center}
\includegraphics[width=0.88\columnwidth]{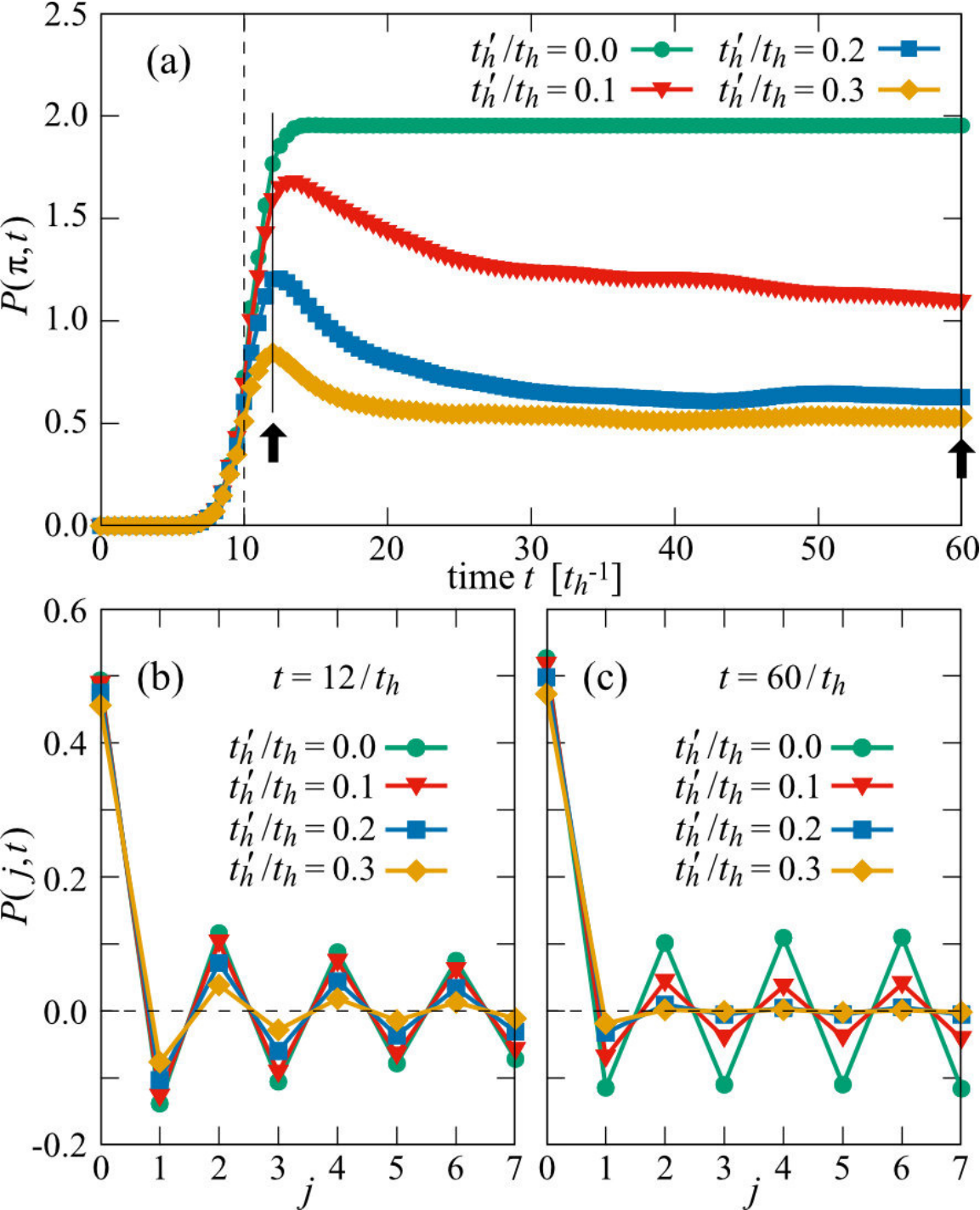}
\caption{
(a)~Time evolution of the pair structure factor $P(q \! = \! \pi,t)$ in the 1D Hubbard model with the NNN hopping $t'_h$ 
at half-filling. 
Two arrows indicate the time, $t=12/t_h$ and $60/t_h$, at which the 
real-space pair correlation function $P(j,t)$ is calculated in (b) and (c), respectively. 
Note that $t=12/t_h$ in (b) is within the transient period. 
The results are calculated by the ED method for $L=14$ (PBC) at $U=8t_h$ with $A_0=0.4$, $\omega_p=8.2t_h$, 
$\sigma_p=2/t_h$, and $t_0=10/t_h$ for the time-dependent vector potential $A(t)$. 
}
\label{s-fig6}
\end{center}
\end{figure}

Figure~\ref{s-fig6} shows the time dependence of the pair correlation functions for the photoexcited state 
$|\Psi(t)\rangle$ with different values of $t'_h$ calculated by the ED method for $L=14$ under PBC. 
As in the main text, the time-dependent external field is introduced via the Peierls phase through 
the time-dependent vector potential $A(t)$, 
where the Peierls phase for the NNN hopping $t'_h$ is given as 
$ t'_h \hat{c}_{j,\sigma}^{\dag}\hat{c}_{j+2,\sigma} \rightarrow t'_h e^{2iA(t)} \hat{c}_{j,\sigma}^{\dag} \hat{c}_{j+2,\sigma}$ 
and the form of $A(t)$ is described in the main text.  

Although the $\eta$ commutation relations are broken when $t'_h$ is finite in $\hat{\cal{H}}_{\rm NNN}$, 
we find the enhancement of the pair correlation functions, specially during the transient period, with   
the $\eta$-pairing like sign-alternating oscillation [see Fig.~\ref{s-fig6}(b)]. 
Note that, unlike in the case of $t'_h=0$, $P(q \! = \! \pi,t)$ is no longer conserved after the pulse irradiation 
because of $[\hat{\mathcal{H}}_{t'_h},\hat{\eta}^{+}\hat{\eta}^{-}] \ne 0$.
With increasing $t'_h$, $P(j ,t)$ at $j>0$  becomes suppressed and eventually show no longer range correlation 
after the pulse irradiation [see Fig.~\ref{s-fig6}(c)]. 
Therefore, we conclude that the photoinduced states still show the robust $\eta$-pairing correlations transiently as long as 
the NNN hopping $t'_h$ is small, although the large NNN hopping $t'_h$ is unfavorable for the photoinduced $\eta$-pairing. 

\begin{figure}[thb]
\begin{center}
\includegraphics[width=0.88\columnwidth]{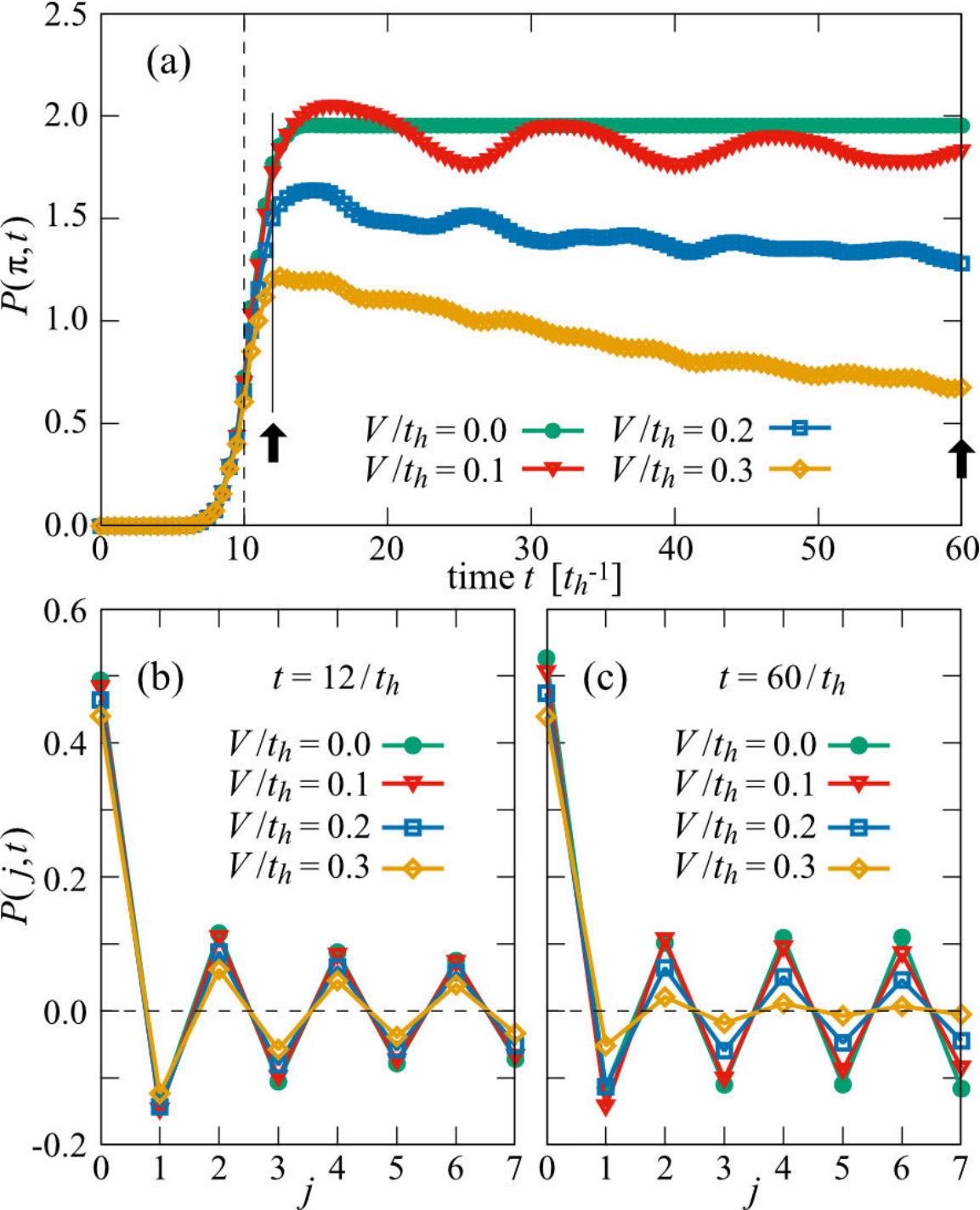}
\caption{
(a)~Time evolution of the pair structure factor $P(q \! = \! \pi,t)$ in the 1D Hubbard model with the 
nearest-neighbor Coulomb interaction $V$ at half-filling. 
Two arrows indicate the time, $t=12/t_h$ and $60/t_h$, at which the real-space 
pair correlation function $P(j,t)$ is calculated in (b) and (c), respectively. 
Note that $t=12/t_h$ in (b) is within the transient period. 
The results are calculated by the ED method for $L=14$ (PBC) at $U=8t_h$ with $A_0=0.4$, 
$\omega_p=8.2t_h$, $\sigma_p=2/t_h$, and $t_0=10/t_h$ for the time-dependent vector potential $A(t)$. 
}
\label{s-fig7}
\end{center}
\end{figure}


\subsection{1D Hubbard model with the nearest-neighbor Coulomb interaction} 

In addition, we examine the influence of the nearest-neighbor Coulomb interaction $V$ on the pair correlation 
in the photoinduced state. 
The model considered here is the 1D extended Hubbard model described by 
$\hat{\cal{H}}_{\rm EH}=\hat{\cal{H}} + \hat{\mathcal{H}}_{V} $, where 
the intersite Coulomb interaction term is given as  
\begin{equation}
\hat{\mathcal{H}}_{V} = V \sum_{j,\sigma,\sigma'}  \hat{n}_{j,\sigma} \hat{n}_{j+1,\sigma'}. 
\end{equation}
Because $[\hat{\mathcal{H}}_{V},\hat{\eta}^{+}] = 2V \sum_{j,\sigma} (-1)^j \hat{c}^{\dag}_{j,\downarrow} \hat{c}^{\dag}_{j,\uparrow} (\hat{n}_{j-1,\sigma} + \hat{n}_{j+1,\sigma})$, the Hamiltonian $\hat{\cal{H}}_{\rm EH}$ 
breaks the $\eta$ commutation relations. 
Figure~\ref{s-fig7} shows the time dependence of the pair correlation functions 
for the photoexcited state $|\Psi(t)\rangle$ with different values of $V$ 
calculated by the ED method for $L=14$ 
under PBC. The time-dependent external field is introduced exactly in the same form described in the main text. 

Although the $\eta$ commutation relations are broken when $V$ is finite in $\hat{\cal{H}}_{\rm EH}$, 
we find the enhancement of the pair correlation functions at least in the transient period of the pulse irradiation, 
clearly exhibiting the $\eta$-pairing like sign-alternating oscillation  [see Fig.~\ref{s-fig7}(b)]. 
However, when $V$ is relatively large, the pair correlation is quickly suppressed with increasing $t$ after the 
pulse irradiation [see Fig.~\ref{s-fig7}(c)]. 
Therefore, we conclude that the photoexcited state can still show the robust pair correlation in the transient period, 
but the strong intersite Coulomb interaction $V$ can eventually disturb the photoinduced $\eta$-pairing completely 
for large $t$.


\subsection{Other related studies for $\eta$-pairing}

In nonequilibrium contexts, possible realization of the $\eta$-paring state has also been proposed in the 
repulsive Hubbard systems with the harmonic trapping potential~\citeSM{RRBetal08S} 
and with the dissipative coupling~\citeSM{BBPetal13S}. 
However, unlike these studies, our mechanism shown here is based on the selection rule derived from 
the commutation relation between the $\eta$ pair and current $\hat{J}$ operators, and therefore 
provides a completely different pathway of $\eta$ pair generation. 

In the attractive Hubbard model, Kitamura and Aoki have investigated the $\eta$-pairing state induced by the periodically 
driven field~\citeSM{KA16S}. 
Based on the Floquet formalism for the effective model in the strong coupling limit, 
composed of the pair hopping term and the nearest-neighbor pair repulsion, 
which can be mapped onto a Heisenberg like model~\citeSM{RSB90S}, 
they have shown that 
the $\eta$-paring state can be induced from the $s$-wave superconducting state 
by varying the effective model parameters~\citeSM{KA16S}. 
However, the corresponding argument cannot be applied to the repulsive 
Hubbard model~\citeSM{MBE15S},  
and thus our mechanism also differs from their suggestion.

\bibliographystyleSM{apsrev4-1}
\bibliographySM{TE-PI_Paper}

\end{document}